\documentclass[sigconf]{acmart}

\usepackage{epsfig,endnotes}
\usepackage{booktabs}
\usepackage{multirow}
\usepackage{rotating}
\usepackage{listings}
\usepackage{upquote}
\usepackage{amssymb}
\usepackage{pifont}
\usepackage{amsfonts}
\usepackage{extarrows}
\usepackage{algorithm}
\usepackage{algpseudocode}
\usepackage{array}
\usepackage{tikz}
\usepackage{xurl}
\usepackage{multirow}
\usepackage{times}
\usepackage{colortbl}
\usepackage{subcaption}
\usepackage{cleveref}
\usepackage{graphicx}
\usepackage{epstopdf}
\usepackage[colorinlistoftodos]{todonotes}
\usepackage{makecell}
\usepackage{textcomp}
\usepackage{diagbox}

\usepackage{enumitem}

\begin{document}

\newcommand{\myname}{{\sc CryptoGuard}}

\newcommand{\bench}{{\sc CryptoApi-Bench}}

\title{\myname{}: High Precision Detection of Cryptographic Vulnerabilities in Massive-sized Java Projects}

\newcommand\Mark[1]{\textsuperscript#1}

\author{Sazzadur Rahaman\Mark{1}, Ya Xiao\Mark{1}, Sharmin Afrose\Mark{1}, Fahad Shaon\Mark{2}, Ke Tian\Mark{1}, Miles Frantz\Mark{1}, \\ Danfeng (Daphne) Yao\Mark{1}, Murat Kantarcioglu\Mark{2}}
\affiliation{ 
     \institution{\Mark{1}Computer Science, Virginia Tech, Blacksburg, VA \\
      \Mark{2}Computer Science, University of Texas at Dallas, Dallas, TX}
    }
\email{{sazzad14, yax99, sharminafrose, ketian, frantzme, danfeng}@vt.edu, {Fahad.Shaon, muratk}@utdallas.edu}

%


\newcommand{\etal}{\emph{et al. }}
\newcommand{\eg}{\emph{e.g.}}
\newcommand{\ie}{\emph{i.e.}}

\newcommand{\apachealerts}{1,295}
\newcommand{\apachetruep}{1,277}
\newcommand{\precision}{98.61\%}
\newcommand{\apachereduction}{76\%}
\newcommand{\androidreduction}{80\%}
\newcommand{\xmark}{\ding{55}}

\let\origthelstnumber\thelstnumber
\settopmatter{printacmref=false}
\renewcommand\footnotetextcopyrightpermission[1]{}

\makeatletter

\newcommand\xsmall{\@setfontsize\small{7}{8}}
\renewcommand\scriptsize{\@setfontsize\scriptsize{6}{7}}
\renewcommand{\textrightarrow}{$\rightarrow$}

\renewcommand{\shortauthors}{}

\newcommand{\hyph}{-\penalty\z@\hskip\z@skip }
\newcommand{\removelatexerror}{\let\@latex@error\@gobble}

\newcommand*\Suppressnumber{%
  \lst@AddToHook{OnNewLine}{%
    \let\thelstnumber\relax%
     \advance\c@lstnumber-\@ne\relax%
    }%
}

\newcommand*\circled[1]{
  \tikz[baseline=(X.base)] 
  \node (X) [draw, shape=circle, inner sep=0pt, fill=white, text=black] {\strut #1};
}

\newcommand*\Reactivatenumber[1]{%
  \setcounter{lstnumber}{\numexpr#1-1\relax}
  \lst@AddToHook{OnNewLine}{%
   \let\thelstnumber\origthelstnumber%
   \refstepcounter{lstnumber}
  }%
}
\makeatother

\begin{abstract}
Cryptographic API misuses, such as exposed secrets, predictable random numbers, and vulnerable certificate verification, seriously threaten software security. The vision of automatically screening cryptographic API calls in massive-sized (e.g., millions of LoC) Java programs is not new. However, hindered by the practical difficulty of reducing false positives without compromising analysis quality, this goal has not been accomplished. State-of-the-art crypto API screening solutions are not designed to operate on a large scale. 

Our technical innovation is a set of fast and highly accurate slicing algorithms. Our algorithms refine program slices by identifying language-specific irrelevant elements. The refinements reduce false alerts by 76\% to 80\% in our experiments. Running our tool, \myname{}, on 46 high-impact large-scale Apache projects and 6,181 Android apps generate many security insights. Our findings helped multiple popular Apache projects to harden their code, including Spark, Ranger, and Ofbiz. We also have made substantial progress towards the science of analysis in this space, including:
{\em i)} manually analyzing \apachealerts{} Apache alerts and confirming \apachetruep{} true positives (\precision{} precision), 
{\em ii)} creating a benchmark with 38-unit basic cases and 74-unit advanced cases, 
{\em iii)} performing an in-depth comparison with leading solutions including CrySL, SpotBugs, and Coverity.  We are in the process of integrating \myname{} with the Software Assurance Marketplace (SWAMP).


\end{abstract}
\sloppy

%
%


\keywords{accuracy, cryptographic API misuses, static program analysis, false positive, benchmark}

\maketitle

\section{Introduction}


Cryptographic algorithms offer provable security guarantees in the presence of adversaries. However, vulnerabilities and deficiencies in low-level cryptographic implementations seriously reduce the guarantees in practice~\cite{boneh46twenty, DBLP:conf/ccs/AdrianBDGGHHSTV15, DBLP:conf/uss/Garman0KMR16, DBLP:conf/ccs/CheckowayMGFC0H16, DBLP:conf/ccs/GarciaBY16}. 
Researchers also found misusing cryptographic APIs is not unusual in application-level code~\cite{DBLP:conf/ccs/EgeleBFK13}. 
Causes of these vulnerabilities are multi-fold, which include complex APIs~\cite{SP-crypto-API-2017, Bodden-crypto-API-2015}, the lack of cybersecurity training~\cite{Meng-ICSE-2018}, the lack of tools~\cite{Survey-SecDev-2017}, and insecure and misleading forum posts (such as on StackOverflow)~\cite{DBLP:conf/sp/AcarBFKMS16,Meng-ICSE-2018}. Some aspects of security libraries (such as JCA, JCE, and JSSE\footnote{JCA, JCE, and JSSE stand for Java Cryptography Architecture, Java Cryptography Extension, and Java Secure Socket Extension, respectively.}) are difficult for developers to use correctly, e.g., certificate verification~\cite{DBLP:conf/ccs/GeorgievIJABS12} and cross-language encryption and decryption~\cite{Meng-ICSE-2018}.



In this work, we focus on the goal of screening massive-sized Java projects for cryptographic API misuses. Specifically, we aim to design a static analysis tool that has no or few false positives (i.e., false alarms) and can be routinely used by developers.

Efforts to screen cryptographic APIs have been previously reported in the literature, including static analysis (e.g., CrySL~\cite{DBLP:conf/ecoop/KrugerS0BM18}, FixDroid~\cite{DBLP:conf/ccs/NguyenWABWF17},
CogniCrypt~\cite{DBLP:conf/kbse/KrugerNRAMBGGWD17},
CryptoLint~\cite{DBLP:conf/ccs/EgeleBFK13}) and dynamic analysis (e.g., SMV-Hunter~\cite{DBLP:conf/ndss/SounthirarajSGLK14}, and AndroSSL~\cite{DBLP:conf/fps/GagnonFFDOB15}), as well as manual code inspection~\cite{DBLP:conf/ccs/GeorgievIJABS12}. Static and dynamic analyses have their respective pros and cons. Static methods do not require the execution of programs. They scale up to a large number of programs, cover a wide range of security rules, and are unlikely to have false negatives (i.e., missed detections). 
Dynamic methods, in comparison, require one to trigger and detect specific misuse symptoms at runtime (e.g., misconfigurations of SSL/TLS). The advantage of dynamic approaches is that they tend to produce fewer false positives (i.e., false alarms) than static analysis. Deployment-grade code screening tools need to be scalable with wide coverage. Thus, static program analysis approach is favorable. 
However, existing static analysis-based tools (e.g.,~\cite{DBLP:conf/ecoop/KrugerS0BM18,DBLP:conf/ccs/NguyenWABWF17,DBLP:conf/kbse/KrugerNRAMBGGWD17,DBLP:conf/ccs/EgeleBFK13}) are not optimized to operate on the scale of massive-sized Java projects (e.g., millions of LoC), which we explain later. 
Existing static analysis tools are also limited in detecting SSL/TLS API misuses and are not designed to detect complex misuse scenarios.
For example, MalloDroid~\cite{DBLP:conf/ccs/FahlHMSBF12} uses a list of known insecure implementations of \texttt{HostnameVerifier} and \texttt{TrustManager} to screen apps. Google Play recently deployed an automatic app checking mechanism for SSL/TLS hostname verifier and certificate verification vulnerabilities~\cite{Google-auto-checking-1}. However, the inspection appears to only target obvious misuse scenarios, e.g., \verb1return true1 in \verb1verify1 method or an empty body in \verb1checkServerTrusted1~\cite{google-auto-checking-bypass}. 

We made substantial progress toward building a high accuracy and low runtime static analysis solution for detecting cryptographic and SSL/TLS API misuse vulnerabilities. Our tool,\myname{}, is built on specialized forward and backward program slicing techniques. These slicing algorithms are implemented by using flow-, context- and field-sensitive data-flow analysis. 

Although program slicing is a well-known technique for identifying the set of instructions that influence or are influenced by a program variable, its direct application to screening cryptographic implementations has several problems, which are explained next.

{\em Detection accuracy.} A challenging problem that has not been solved by prior work is the excessive number of false positives that basic static analysis (including slicing) generates. Several types of detection require one to search for constants or values from predictable APIs, e.g., passwords, seeds, or initialization vectors (IVs). However, benign constants or irrelevant parameters may be mistaken as violations (e.g., array/collection bookkeeping constants). 
Another source of detection inaccuracy comes from the assumption that all the system and runtime libraries are present during the analysis. This assumption holds for Android apps (e.g., CrySL~\cite{DBLP:conf/ecoop/KrugerS0BM18}, CryptoLint~\cite{DBLP:conf/ccs/EgeleBFK13}), but not necessarily for Java projects.

A feature of our solution \myname{} is a set of refinement algorithms that systematically discard false alerts. These refinement insights are derived from empirical observations of common programming idioms and language restrictions to remove irrelevant resource identifiers, arguments about states of operations, constants on infeasible paths, and bookkeeping values. For eight of our rules, these refinement algorithms reduce the total number of alerts by \apachereduction{} in Apache and \androidreduction{} in Android (Figure~\ref{before:after}). Our manual analysis shows that \myname{} has a precision of \precision{} on Apache.

{\em Runtime overhead and coverage.} Existing flow-, context- and field-sensitive analysis techniques build a super control-flow graph of the entire program, which has a significant impact on runtime. In contrast, our on-demand slicing algorithms run much faster, which start from the slicing criteria and only propagate to the methods that have the potential to impact security. Hence, a large portion of the code base is not touched. 
For the Apache projects we evaluated, \myname{} took around 3.3 minutes on average. 

More importantly, our lightweight analysis building blocks enable us to address complex API misuse scenarios. \myname{} covers more cryptographic properties than CrySL~\cite{DBLP:conf/ecoop/KrugerS0BM18}, Coverity~\cite{coverity}, and SpotBugs~\cite{spotbugs} combined. Our most complex analysis (for Rule 15 on insecure RSA/ECC key sizes) involves multiple rounds of forward and backward slicing.


Our technical contributions are summarized as follows.

\begin{itemize}[leftmargin=2em]

\item
We designed and implemented a set of new analysis algorithms for detecting cryptographic and SSL/TLS API misuses. Our static code checking tool, \myname{}, is designed for developers to use routinely on large Java projects.  Besides open-sourcing \myname{}\footnote{Available at \url{https://github.com/CryptoGuardOSS/cryptoguard} under GPL v3.0.}, we are currently integrating it with the Software Assurance Marketplace (SWAMP)~\cite{swamp}, a well-known free software security analysis platform.

\item 
We gained numerous security insights from screening 46 Apache projects. For $15$ of our rules, we observed violations in Apache projects (Table~\ref{overview:table}). 39 out of the 46 projects have at least one type of cryptographic misuses, and 33 projects have at least two. 
We reported our security findings to Apache, some of which have been promptly fixed. In Section~\ref{sec:discussion}, we share our experience of interacting with Apache teams and their pragmatic constraints e.g., backward compatibility, operation in humanless settings.

\item
Our evaluation on 6,181 Android apps shows that around 95\% of the total vulnerabilities come from libraries that are packaged with the application code. Some libraries are from Google, Facebook, Apache, Umeng, and Tencent (Table~\ref{table:violation-by-packagename}). We observe violations in most of the categories, including hardcoded keyStore passwords, e.g., \verb1notasecret1 is used in multiple Google libraries (Table~\ref{table:app-violation-source-percent}). 
We also detected multiple SSL/TLS (MitM) vulnerabilities that Google Play's automatic screening seemed to have missed. 

\item
We created a benchmark named \bench{} with 112 unit test cases.\footnote{Our benchmark is available at \url{https://github.com/CryptoGuardOSS/cryptoapi-bench}.} \bench{} contains basic intra-procedural instances, inter-procedural cases,  field sensitive cases, false positive tests, and correct API uses. Our evaluation on \bench{} shows that \myname{} achieves higher precision and recall than Coverity, SpotBugs and CrySL~\cite{DBLP:conf/ecoop/KrugerS0BM18}, which is the state-of-the-art research solution. The benchmark also reveals false negatives that \myname{} needs to improve on in the future. 

\end{itemize}

Our key technical novelty and significance are summarized as follows.
{\bf [Formulation of problems]} We present the mappings between a number of cryptographic abstractions to concrete Java programming elements that can be statically enforced. The mapping strategy (including specific slicing criteria) is useful beyond \myname{} (in Section~\ref{mapping-vul-to-pgm}). 
{\bf [Methodology development]} We specialize program slicing with new language-based contextual refinement algorithms and successfully show a significant reduction of false alarms (related to constants and predictable values). It is a substantial advancement over general-purpose slicing and state-of-the-art solutions (in Section~\ref{sec:heuristics}).
{\bf [New security capabilities]}  Our lightweight algorithm design enables \myname{} to check more rules than existing solutions, while maintaining high precision. 
{\bf [New security findings]} \myname{} enables us to report a number of alarming cryptographic coding issues in open source Apache projects and Android (in Sections~\ref{analysis-of-apache} and~\ref{analysis-of-android}).
{\bf [Science of security]} Our \bench{} will motivate researchers to improve the accuracy, coverage, scalability, and transparency of their tools, collectively advancing the science of security (in Section~\ref{sec:hestia-fixdroid}).



%

\begin{figure*}
\begin{minipage}[c]{.65\textwidth}
\begin{minipage}[c]{.52\textwidth}
\lstset{language=Java, 
   basicstyle=\ttfamily\scriptsize,
   keywordstyle=\color{blue}\ttfamily,
   stringstyle=\color{red}\ttfamily,
   commentstyle=\color{cyan}\ttfamily,
   morecomment=[l][\color{magenta}]{\#},
   numbers=left,
   numbersep=1pt,
   escapeinside=||,
   tabsize=1,
   stepnumber=1,
   showstringspaces=false,
   xleftmargin=0pt
}
\begin{lstlisting}
class PasswordEncryptor {

 Crypto crypto;

 public PasswordEncryptor(){
  String passKey = PasswordEncryptor|\Suppressnumber|
                         .getKey(|\underline{"pass.key"}|);|\Reactivatenumber{7}|
   crypto = new Crypto(passKey);|\circled{p}|
 }

 byte[] encPass(String [] arg){
  return crypto.encrypt(arg[0], arg[1]);|\circled{p}|
 }

 static String getKey(String src){
  String key = Context.getProperty(src);
  if (key == null) {
    key = "defaultkey";
  }
  return key;
  }
}
\end{lstlisting}
\end{minipage}
\begin{minipage}[c]{.40\textwidth}
\lstset{language=Java, 
   basicstyle=\ttfamily\scriptsize,
   keywordstyle=\color{blue}\ttfamily,
   stringstyle=\color{red}\ttfamily,
   commentstyle=\color{green}\ttfamily,
   morecomment=[l][\color{magenta}]{\#},
   numbers=left,
   numbersep=1pt,
   escapeinside=||,
   tabsize=1,
   firstnumber=22,
   stepnumber=1,
   showstringspaces=false,
   xleftmargin=0pt
}
\begin{lstlisting}
class Crypto {

 String ALGO = "AES";
 String ALGO_SPEC = "AES/CBC/NoPadding";
 String defaultKey;
 Cipher cipher;

 public Crypto(String defKey){
  cipher = Cipher.getInstance(ALGO_SPEC);
  defaultKey = defKey; // assigning field
 }

 byte[] encrypt(String txt,String key){
  if (key == null){
   key = defaultKey;|\circled{f}|
  }
  byte[] keyBytes = key.getBytes(|\underline{"UTF-8"}|);
  byte[] txtBytes = txt.getBytes();
  SecretKeySpec keySpc =|\Suppressnumber|
    |\Reactivatenumber{41}|new SecretKeySpec(|\underline{keyBytes}|, ALGO);
  cipher.init(Cipher.ENCRYPT_MODE,keySpc);   
  return cipher.doFinal(txtBytes);}}
\end{lstlisting}
\end{minipage}\caption*{(a)}
\end{minipage}
\begin{minipage}[c]{.30\textwidth}
\begin{subfigure}{0.76\textwidth}
 		\includegraphics[width=1.0\textwidth]{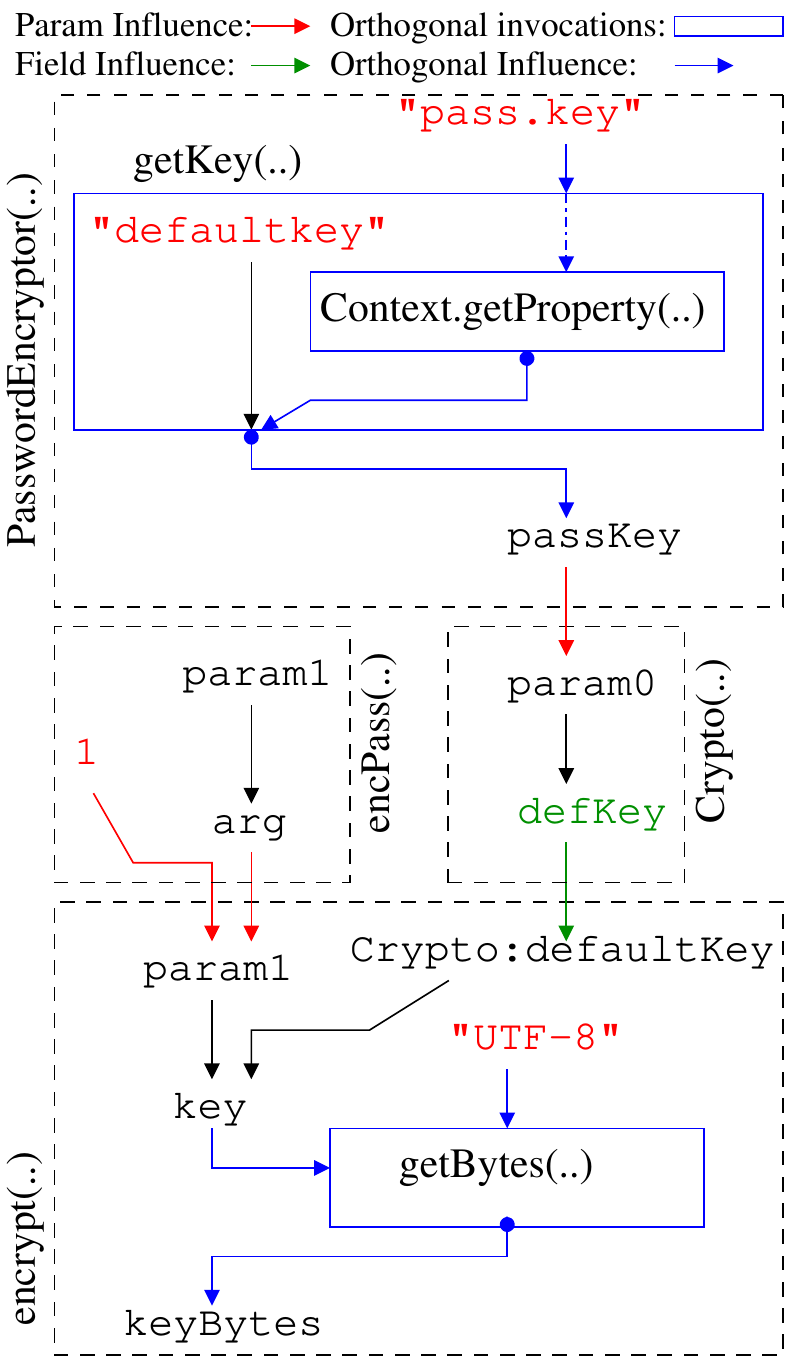}
 		\caption*{(b)}
        \label{program-slicing:call-graph}
 	\end{subfigure} 
\end{minipage}
\caption{(a) An example demonstrating various features of \myname{}. \texttt{Crypto} class is used for generic AES encryption and \texttt{PasswordEncryptor} class uses \texttt{Crypto} for password encryption. \protect\circled{f} indicates influence through the fields and \protect\circled{p} indicates influence through the method parameters. (b) Partial data dependency graph for \texttt{keyBytes} variable.}\label{coding:example}
\vspace{-10px}
\end{figure*}

\section{Threat Model, Challenges, and Overview}
\label{sec:2}
We describe our threat model and discuss the technical challenges associated with detecting these threats with static program analysis. For each challenge, we briefly overview our solution.

\subsection{Threat Model}\label{sec:vulnerabilities}

We summarize the vulnerabilities that \myname{} aims to detect below and in Table~\ref{table:rules}. We also rank their severity.

\noindent
{\em \textbf{1. Vulnerabilities due to predictable secrets.}} 
Software with predictable cryptographic keys and passwords are inherently insecure~\cite{DBLP:conf/ccs/EgeleBFK13}.  
Here, we consider the use of any constants, as well as values that are derived from constants or API calls with predictable outputs (e.g., DeviceID, Timestamps) to be insecure.

\noindent
{\em \textbf{2. Vulnerabilities from MitM attacks on SSL/TLS.}}
Improper customization of Java Secure Socket Extension (JSSE) APIs may result in man-in-the-middle (MitM) vulnerabilities~\cite{DBLP:conf/ccs/FahlHMSBF12,DBLP:conf/ccs/GeorgievIJABS12}. 
CryptoLint~\cite{DBLP:conf/ccs/EgeleBFK13} does not detect these vulnerabilities.



\noindent
{\em \textbf{3. Vulnerabilities from predictable PRNGs.}} 
The predictability of pseudorandom number generators (PRNGs) has been a major source of vulnerabilities~\cite{goldberg1996randomness, DBLP:conf/asiacrypt/BernsteinCCCHLS13, DBLP:conf/uss/HeningerDWH12}.
The use of \texttt{java.util.Random} as a PRNG is insecure~\cite{java:util/random, DBLP:conf/crypto/Krawczyk89}.  
In addition, seeds for \texttt{java.security.SecureRandom}~\cite{java:util/securerandom} should not be predictable. 

\noindent
{\em \textbf{4. Vulnerabilities from CPA.}} Ciphertexts should be indistinguishable under chosen plaintext attacks (CPA)~\cite{DBLP:conf/ccs/EgeleBFK13}. Static salts make dictionary attacks easier on password-based encryption (PBE). In addition, static initialization vectors (IVs) in  cipher block chaining (CBC) and electronic codebook (ECB) modes are insecure~\cite{DBLP:journals/iacr/Bard04, DBLP:conf/apsys/LazarCWZ14}.


\noindent
{\em \textbf{5. Vulnerabilities from feasible bruteforce attacks.}}
MD5 and SHA1 are susceptible to hash collision~\cite{DBLP:conf/eurocrypt/StevensLW07,DBLP:conf/crypto/StevensBKAM17} and pre-image~\cite{DBLP:journals/iacr/ChangJMS17, hash:rainbow/lifetime} attacks. 
In addition, bruteforce attacks are feasible for 64-bit symmetric ciphers (e.g., DES, 3DES, IDEA, Blowfish)~\cite{DBLP:conf/ccs/BhargavanL16}. 1024-bit RSA/DSA/DH and 160-bit ECC are also weak~\cite{nist:keysize/reco}. RFC 8018 recommends at least 1000 iterations for PBE~\cite{moriarty2017pkcs}.



\noindent
{\em How severe are these vulnerabilities?} Each case has specific attack scenarios documented in the literature. To prioritize alerts, we categorize their severity into high, medium, and low, based on {\em i)} attacker's gain and {\em ii)} attack difficulty.  Vulnerabilities from predictable secrets and SSL/TLS MitM are immediately exploitable and substantially benefit attackers. In Android, an application can only access its own KeyStore. Hence, hard-coded passwords are less harmful in Android. However, privilege escalation attacks bypass this restriction, which has been demonstrated~\cite{DBLP:conf/ccs/VeenFLGMVBRG16}. Commercially available rainbow tables allow attackers to easily obtain pre-images of MD5 and SHA1 hashes for typical passwords~\cite{hash:rainbow/tables}. Hash collisions for these algorithms enable attackers to forge digital signatures or break the integrity of any messages~\cite{DBLP:conf/crypto/StevensBKAM17, DBLP:conf/ndss/BhargavanL16}. 
Therefore, these vulnerabilities are classified as high risks. Vulnerabilities from predictability and CPA provide substantial advantages to attackers by significantly reducing attack efforts. They are medium-level risks. Brute-forcing ciphers, requiring non-trivial effort, is low risk.

\subsection{Technical Challenges and Solution Overview}
\label{sec:challenge-I}

The task of screening millions of lines of code for cryptographic API misuses poses a set of technical challenges. 

\vspace{5pt}
\noindent
{\em Technical Challenge I: False positives.}

\vspace{2pt}
\noindent
{\em 1. False positives due to phantom methods.}
A method is phantom if its body is not available during analysis.
Unlike Android, Java web applications have phantom libraries. A non-system library that is not packaged with the project binaries is referred to as a phantom library. Existing cryptographic misuse vulnerability solutions (e.g., CryptoLint~\cite{DBLP:conf/ccs/EgeleBFK13}, CrySL~\cite{DBLP:conf/ecoop/KrugerS0BM18}) are not designed to handle phantom libraries, which may cause false positives. For example, in Figure~\ref{coding:example}(a) if the class \texttt{Context} is a member of a phantom library, then \texttt{getProperty} method (Line 15) is a phantom method. The data-flow diagram in Figure~\ref{coding:example}(b) shows that a straightforward def-use analysis would likely report \texttt{pass.key} as a hard-coded key, since it cannot explore \texttt{getProperty} method at Line 15. 

Our solution is a set of new algorithms to refine slicing outputs (Section~\ref{sec:heuristics}). For example, examining the context reveals that \texttt{pass.key} is used as an identifier of a key and has no security influence on \texttt{keyBytes}. Thus, it can be safely discarded. 

\vspace{2pt}
\noindent
{\em 2. False positives due to data structures.}
 Constants for bookkeeping data structures are another major source of false positives that are largely uncovered in the existing literature (e.g.,~\cite{DBLP:conf/ecoop/KrugerS0BM18,DBLP:conf/ccs/EgeleBFK13}). Most frequently used data structures include lists, maps, and arrays. For example, a data-structure-unaware analysis would likely report \texttt{1} from Line 11 (Figure~\ref{coding:example}(a)) as a hard-coded key, as it influences the \texttt{key} parameter of \texttt{encrypt} method (Figure~\ref{coding:example}(b)).  Our refinement algorithms track and discard any kinds of data-structure-bookkeeping constants (Section~\ref{sec:heuristics}).

\vspace{5pt}
\noindent
{\em Technical Challenge II: precision vs. runtime tradeoff.}
For a project with millions LoC (e.g., Apache Hadoop has 2.5 million LoC), building a super-CFG is costly and unnecessary. Cryptographic functionality is often confined within a small fraction of the project. However, most flow-, context- and field-sensitive analysis based tools (e.g.,~\cite{DBLP:conf/ccs/EgeleBFK13,DBLP:conf/ecoop/KrugerS0BM18}) appear to build a super control-flow graph, e.g., by superimposing the project's call graph over control-flow graphs of methods, adding call edges between
\textit{invoke} instructions, method entries, and exits.

In contrast, we adopt the following more scalable approaches.

\begin{table*}[]
\centering
\caption{Cryptographic vulnerabilities, properties, and static analysis methods used. High, medium, and low risk levels are denoted by H/M/L, respectively. CPA stands for chosen ciphertext attack, MitM for man-in-the-middle, C/I/A  for confidentiality, integrity, and authenticity, respectively.  $\uparrow$ means backward slicing and $\downarrow$ means forward slicing. Slicing is inter-procedural unless otherwise specified (e.g., intra, both). Refinement insights are applied for all the inter-procedural backward slicing.}
\label{table:rules}
\resizebox{\textwidth}{!}{%
\begin{tabular}{@{}lllllll@{}}
\toprule
\textbf{No} & \textbf{Vulnerabilities} & \textbf{Attack Type} & \textbf{Crypto Property} & \textbf{Severity} & \textbf{Our Analysis Method} \\ \midrule
1 & Predictable/constant cryptographic keys. & \multirow{3}{*}{Predictable Secrets} & Confidentiality & H & $\uparrow$ slicing \& $\downarrow$ slicing \\
2 & Predictable/constant passwords for PBE &  & Confidentiality & H & $\uparrow$ slicing \& $\downarrow$ slicing\\
3 & Predictable/constant passwords for KeyStore &  & Confidentiality & H & $\uparrow$ slicing \& $\downarrow$ slicing\\
\hline
4 & Custom Hostname verifiers to accept all hosts & \multirow{4}{*}{SSL/TLS MitM} & C/I/A & H & $\uparrow$ slicing (intra) \\
5 & Custom TrustManager to trust all certificates &  & C/I/A & H & $\uparrow$ slicing (intra)\\
6 & Custom SSLSocketFactory w/o manual Hostname verification &  & C/I/A & H & $\downarrow$ slicing (intra)\\
7 & Occasional use of HTTP &  & C/I/A & H & $\uparrow$ slicing\\
\hline
8 & Predictable/constant PRNG seeds & \multirow{2}{*}{Predictability} & Randomness & M & $\uparrow$ slicing \& $\downarrow$ slicing \\
9 & Cryptographically insecure PRNGs (e.g., java.util.Random) &  & Randomness & M & Search\\
\hline
10 & Static Salts in PBE & \multirow{3}{*}{CPA} & Confidentiality & M & $\uparrow$ slicing \& $\downarrow$ slicing\\
11 & ECB mode in symmetric ciphers &  & Confidentiality & M & $\uparrow$ slicing\\
12 & Static IVs in CBC mode symmetric ciphers &  & Confidentiality & M & $\uparrow$ slicing \& $\downarrow$ slicing\\
\hline
13 & Fewer than 1,000 iterations for PBE & \multirow{4}{*}{Brute-force} & Confidentiality & L & $\uparrow$ slicing \& $\downarrow$ slicing\\
14 & 64-bit block ciphers (e.g., DES, IDEA, Blowfish, RC4, RC2) &  & Confidentiality & L & $\uparrow$ slicing\\
15 & Insecure asymmetric ciphers (e.g, RSA, ECC) &  & C/A & L & $\uparrow$ slicing \& $\downarrow$ slicing (both)\\
16 & Insecure cryptographic hash (e.g., SHA1, MD5, MD4, MD2) &  & Integrity & H & $\uparrow$ slicing \\ \cmidrule(){1-6} 
\end{tabular}
}%
\end{table*}

\vspace{2pt}
\noindent
{\em 1. Demand-driven analysis.}
Our flow- and context- sensitive analysis is demand driven. Initially, it only creates a call graph. During the analysis, it performs on-demand inter-procedural backward data flow analysis to perform backward slicing where the analysis starts from the slicing criteria and propagates {\em upward} and {\em orthogonally} on-demand. For example, in Figure~\ref{coding:example}(a), a propagation from \texttt{encrypt} method to \texttt{encPass} method, is an upward propagation. A propagation to orthogonal method invocations at Line 6 and 38 are orthogonal propagation.
Our field sensitivity is also demand driven. Field sensitivity is applied to a field if the field is directly or indirectly used in our inter-procedural backward slices. A field's influence is considered indirect if the field is accessed using orthogonal method invocations (i.e., getter methods). We refer to this field sensitivity as \textit{data-only class field-sensitivity.} 

\vspace{2pt}
\noindent
{\em 2. Control the depth of orthogonal explorations.} Most of our cryptographic vulnerabilities involve finding constants. A distinguishing feature of constants is that they require no or few processing before use. Generally, processing is done by orthogonal method invocations. Thus, we explore the trade-off between 
the depth of orthogonal explorations and the runtime/accuracy of the analysis, by clipping orthogonal explorations up to a predetermined level with a low likelihood of causing false negatives. (We set the depth to 1 in our experiments.) 
We then use similar techniques as in phantom methods handling to reduce the false positives introduced by clipping. 

\vspace{2pt}
\noindent
{\em 3. Subproject awareness.} Code in large Java projects is usually organized into reusable subprojects, packaged as separate \verb1.jar1s. One may create a \textit{fat jar} by including all subprojects to resolve inter-subproject dependencies, which would unnecessarily increase the call graph generation time and size. In contrast, \myname{} creates and consults a directed acyclic graph (DAG) representing subproject dependencies. It helps {\em i)} exclude unnecessary subprojects and {\em ii)} analyze independent sub-projects concurrently.


\section{Mapping Vulnerabilities to Program Analysis}
\label{mapping-vul-to-pgm}

It is important to map cryptographic properties to concrete Java programming elements that can be statically enforced. We break the detection plan into one or more abstract steps so that each step can be mapped to a single round of static analysis.


In this section, we illustrate the process of mapping cryptographic vulnerabilities to concrete program analysis tasks. This mapping process is manual and only needs to be performed once for each vulnerability. In what follows, we use rule $i$ to refer to the detection of vulnerability $i$ in Table~\ref{table:rules}. 

For example, in Rule 4 we detect the abuse of \texttt{HostnameVerifier} interface. Ideally, an implementation of \texttt{HostnameVerifier} must use the \texttt{javax.net.ssl.SSLSession} parameter \texttt{verify} method to verify the hostname. Using the \texttt{return} statement as the slicing criterion, we perform intra-procedural backward slicing of \texttt{verify} method to implement this rule.

Rule 5 is to detect the abuse of the \texttt{X509TrustManager} interface. We reduce the task to detecting 3 concrete cases: {\em i)} throwing no exception after validating a certificate in \texttt{checkServerTrusted}, {\em ii)} unpinned self-signed certificate with an expiration check, and {\em iii)} not providing a valid list of certificates in \texttt{getAcceptedIssuers}.
For Case {\em i)}, intuitively, our program analysis needs to search for the occurrences of \texttt{throw} or propagated exception.  \texttt{throw} is the slicing criterion in the (intra-procedural) backward slicing. Simple parsing is inadequate, as the analysis needs to learn the type of the thrown exception. 

Rule 6 is to detect whether any method uses \texttt{SSLSocket} directly without performing hostname verification. Intuitively, to detect this vulnerability, we need to track whether an \texttt{SSLSocket} created from \texttt{SSLSocketFactory} influences the \texttt{SSLSession} parameter of a \texttt{verify} method (of a \texttt{HostnameVerifier}) invocation. In addition, we also need to check whether the return value of the \texttt{verify} method is used in a condition checking statement (e.g., \texttt{if}). For detection, we use forward program slicing to identify all the instructions that are influenced by the \texttt{SSLSocketFactory} instance. Among these instructions, we examine three cases {\em i)} an \texttt{SSLSocket} is created, {\em ii)} an \texttt{SSLSession} is created and used in \texttt{verify}, and {\em iii)} the return value of \texttt{verify} method is used to make decisions. These three cases represent a correct use of \texttt{SSLSocket} with proper hostname verification.


Rule 15 is to detect insecure asymmetric cipher configurations (e.g., 1024-bit RSA). A more concrete goal is to detect an insecure default key size use and an explicit definition of insecure key size. 
The tasks of program analysis are to determine {\em a)} whether the key size is defined explicitly or by default, {\em b)} the statically defined key size, and {\em c)} the key generation algorithm. 
For Task {\em a)}, our analysis uses forward slicing to determine whether the \texttt{initialize} method is invoked to set the key size of a key-pair generator. For Tasks {\em b)} and {\em c)}, we use two rounds of backward program slicing to determine the key size and algorithm, respectively. We also employ on-demand field sensitivity for data-only classes in Task {\em b)}.
The analyses for Rule 15 are the most complex in \myname{}.

Mappings for other rules are relatively straightforward and can be deduced from Table~\ref{table:rules}. For example, $\uparrow$ in Rule 1 \& 2 means these rules are implemented using inter-procedural backward slicing and $\downarrow$ indicates inter-procedural forward slicing is used for on-demand data-only class field sensitivity.
We also list the slicing criteria used for each rule in Tables~\ref{tab:methods-to-slice}, ~\ref{tab:forward-slicing-criteria} and~\ref{tab:inter-slicing-criteria} in Appendix.


\section{Crypto-specific Slicing} 
\label{sec:RIs}

We specialize static def-use analysis~\cite{DBLP:conf/aswec/YangTM08} and forward and backward program slicings~\cite{DBLP:conf/scam/Lucia01} for detecting Java cryptographic API misuses. 
We break the detection strategy into one or more steps, so that a step can be realized with a single round of program slicing. After performing the slicing, each program slice is analyzed to find the presence of a vulnerability. Our 16 categories of vulnerabilities require different program analysis methods for detection. Table~\ref{table:rules} summarizes slicing techniques to detect each of the vulnerabilities.
General-purpose slicing alone is inadequate. Thus,  we explain our solution for overcoming the accuracy challenge in Section 5. 

A definition of variable $v$ is a statement that modifies $v$ (e.g., declaration, assignment). A use of variable $v$ is a statement that reads $v$ (e.g., a method call with $v$ as an argument). Def-use data-flow analysis or def-use analysis identifies the definition and use statements and describes their dependency relations.  
Given a slicing criterion, which is a statement or a variable in a statement (e.g., a parameter of an API), backward program slicing is to compute a set of program statements that affect the slicing criterion in terms of data flow.
Given a slicing criterion, forward program slicing is to compute a set of program statements that are affected by the slicing criterion in terms of data flow. 
Given a program and a slicing criterion, a program slicer returns a list of program slices. Intra-procedural program slicing mechanisms use def-use analysis to compute slices. 

To confine inter-procedural backward slicing within security code regions, the analysis starts from cryptographic APIs and follows their influences recursively. This approach effectively skips the bulk of the functional code and substantially speeds up the analysis.

\subsection{Slicing Criteria and Backward Slicing}

We give the intuition behind selecting slicing criteria and then present our backward slicing techniques. The complete list of our slicing criteria and corresponding APIs are shown in Tables~\ref{tab:methods-to-slice},~\ref{tab:forward-slicing-criteria}, and~\ref{tab:inter-slicing-criteria} in Appendix.

\smallskip
\noindent
{\em Slicing criteria.}
The choice of slicing criterion directly impacts the analysis outcomes. We choose slicing criteria based on several factors, including  relevance to the vulnerability, simplicity of checking rules, shared across multiple projects. 

For inter-procedural backward slicing, the slicing criteria are defined as the parameters of a target method's invocation. For example, to find predictable secrets (in Rules 1-3), we use the key parameter of the constructors of \texttt{SecretKeySpec} as the slicing criterion.
For intra-procedural backward slicing, we define three types of slicing criteria: {\em i)} parameters of a method, {\em ii)} assignments, and {\em iii)} \texttt{throw} and \texttt{return}. For example, to detect insecure hostname verifiers that accept all hosts (in Rule 4), we use the \texttt{return} statement in the \texttt{verify} method as the slicing criterion.

\smallskip
\noindent
{\em Intra-procedural backward slicing.} The purpose of intra-procedural backward slicing is two-fold. It is used independently to enforce security as well as a building block of inter-procedural back program slicing. The intra-procedural program slicing utilizes the def-use property of a statement to decide whether a statement should be included in a slice or not. Our implementation utilizes the worklist algorithm from the intra-procedural data-flow analysis framework of Soot.
During this process, if any orthogonal method invocations are encountered, it recursively slices them to collect the arguments and statements that influence any field or return statements within that orthogonal methods. To reduce runtime overhead, such orthogonal method explorations are clipped at a pre-configurable depth (1 in our experiments). In this procedure, we use refinement insights presented in Section~\ref{sec:heuristics} to exclude security irrelevant instructions that basic use-def analysis cannot identify.
%
%
%
%

\smallskip
\noindent
{\em On-demand Inter-procedural backward slicing.}\label{sec:inter-procedural-bs}
The main responsibility of this algorithm is the upward propagation of the analysis. Our inter-procedural backward slicing builds on intra-procedural backward slicing. 
Major steps of the algorithm are as follows. {\em i)} We build a caller-callee relationship graph of all the methods of the program. The call-graph construction uses class-hierarchy analysis. {\em ii)} We identify all the callsites of the method specified in the slicing criterion. A callsite refers to a method invocation. {\em iii)} For all the callsites, we obtain all the inter-procedural backward slices by invoking intra-procedural slicing recursively to follow the caller chain.
{\em iv)} Our procedure is field sensitive. Typical field initialization statements are assignments. After encountering a field assignment, the analysis follows the influences through fields, recursively. 

\subsection{Forward Slicing}\label{slicing:forward}
Some of our analysis demands forward slicing, which inspects the statements occurring after the slicing criterion. 


\noindent
{\em Intra-procedural forward slicing.}  We design intra-procedural forward slicing for Rules 6 (SSLSocketFactory w/o Hostname verification) and 15 (Weak asymmetric crypto). 
The operation of intra-procedural forward slicing is similar to that of intra-procedural backward slicing. In forward slicing, we choose assignments as the slicing criteria. 
The traversal follows the order of the execution, i.e., going forward.
Because problematic code regions for Rules 6 and 15 are confined within a method, their forward slicing analyses do not need to be inter-procedural. 

\begin{figure}
    \centering
    \begin{verbatim}
$r1.setText("mytext");
$r1.setKey("mykey");
...
key = $r1.getKey();
    \end{verbatim}
    \vspace{-15pt}
    \caption{Indirect field access using orthogonal invocations on \textit{data-only class} object \texttt{\$r1}.}
    \label{fig:dataonly}
\end{figure}

\noindent
{\em Inter-procedural forward slicing.}

Given an assign instruction or a constant as the slicing criterion, we perform the inter-procedural forward slicing to identify the instructions that are influenced by the slicing criterion in terms of def-use relations. Our version of inter-procedural forward slicing operates on the slices obtained from inter-procedural backward program slicing. Our inter-procedural backward slicing produces an ordered collection of instructions combined from all visited methods.

{\em We define a class as a data-only class, if the fields of the class are only visible within orthogonal method invocations.} We use inter-procedural forward slicing for on-demand field sensitivity of data-only classes, as the field sensitivity during upward propagation (inter-procedural backward slicing) does not cover them. In Figure~\ref{fig:dataonly}, \texttt{\$r1} is an object of data-only class, where its fields are accessed indirectly with an orthogonal method (i.e, \textit{getKey}) invocation. 
Given a constant, using inter-procedural forward slicing, \myname{} determines whether the constant influences any field of a data-only class object and records it. Later on, when it encounters an assign invocation on the same object and observes that the previously recorded field influences the return statement, then it reports the constant. Through this on-demand field sensitivity for data-only class, \myname{} knows that constant \texttt{mytext} (Figure~\ref{fig:dataonly}) is not a hard-coded key. $\downarrow$ in Table~\ref{table:rules} represents the use of forward slicing for on-demand data-only class field sensitivity~\footnote{Current prototype uses this field sensitivity for 8 rules.}.


\section{Refinement for FP Reduction}\label{sec:heuristics}

We design a set of refinement algorithms to exclude security irrelevant instructions to reduce false alarms. These \textit{refinement insights (RI)} are deduced by observing common programming idioms and language restrictions. We also discuss the possibility of  false negatives (i.e., missed detection).

\subsection{Overview of Refinement Insights (RI)}

Eight of our rules (1, 2, 3, 8, 10, 12, 13 and 15) require identifying constants/predictable values in a program slice. The purpose is to ensure that no data (e.g., cryptographic keys, passwords, IVs, and seeds) is hardcoded or solely derived from any hardcoded values. Use of any predictable values (e.g., Timestamp, DeviceID) is also insecure for Rules 1, 2, 3 and 8. However, there are many constant/predictable values that do not impact security. We refer to them as \textit{pseudo-influences}. Pseudo-influences are a major source of false positives.

Based on empirical observations of common programming idioms and language restrictions, we invent five strategies to systematically remove irrelevant constants/predictable values from slices and reduce pseudo-influences, which are summarized next. For eight of our rules, these refinement insights yield a \apachereduction{} reduction in total alerts for Apache projects and \androidreduction{} reduction for Android applications (Section~\ref{analysis-false-alarms}). In \myname{}, rule checkers apply these refinements on constants/predictable values in program slices to remove pseudo-influences.

\begin{itemize}[leftmargin=2em]

\item 

{\em RI-I: Removal of state indicators.} We discard constants/predictable values that are used to describe the state of a variable during an orthogonal method invocation.

\item 

{\em RI-II: Removal of resource identifiers.} We discard constants/predictable values that are used as the identifier of a value source during an orthogonal method invocation.

\item 

{\em RI-III: Removal of bookkeeping indices.} We discard constants/predictable values that are used as the index or size of any data structures. Specifically, RI-III discards any influences on i) size parameter of an array or a collection instantiation, ii) indices of an array, iii) indices of a collection.

\item 

{\em RI-IV: Removal of contextually incompatible constants.} We discard constants/predictable values, if their types are incompatible with the analysis context. For example, a boolean variable cannot be used as a key, IV, or salt.

\item 

{\em RI-V: Removal of constants in infeasible paths.} Some constant initializations are updated along the path to the slicing criterion. We need to discard the initializations that do not have a valid path of influence to the criterion.
\end{itemize}

RI-I, RI-II and RI-IV are used to handle the clipping orthogonal method explorations, which can occur due to phantom method invocations or pre-configured clipping at a certain depth. RI-III is used to achieve data structure awareness and RI-V are used to compensate path insensitivity.
The breakdown of the total reduction of false alarms in our experiment shows that RI-II and RI-III are most effective in Apache and Android apps (Figure~\ref{breakdown}). In the next two subsections, we highlight the details of two refinement insights based on removing state indicators and resource identifiers. Details for other RIs can be found in Section~\ref{other:refinements} of Appendix.

\subsection{RI-I: Removal of State Indicators}\label{def:heuristic:I}

Clipping of orthogonal method exploration can cause false positives if the arguments of method is used to describe the state of a variable. Consider \texttt{UTF-8} in Line 38 of Figure~\ref{coding:example}(a). Its Jimple\footnote{Jimple is an intermediate representation (IR) of a Java program.} representation is as follows, where \texttt{\$r2} represents variable \texttt{key}, \texttt{\$r4} represents \texttt{keyBytes}, and \texttt{virtualinvoke} is for invoking the non-static method of a class.
 
\begin{footnotesize}
\texttt{\$r4 = virtualinvoke \$r2.<java.lang.String: byte[] getBytes(java.lang.String)>("UTF-8")}
\end{footnotesize}
 
If the analysis is clipped so that it cannot explore the \texttt{getBytes} method, then a def-use analysis shows that constant \texttt{UTF-8} influences the value of \texttt{\$r4} (i.e., \texttt{keyBytes}). Thus, a straightforward detection method would report \texttt{UTF-8} as a hardcoded key. However, \texttt{UTF-8} is for describing the encoding of \texttt{\$r2} and can be safely ignored. We refer to this type of constants as \textit{state indicator pseudo-influence}.
 
The use of refinement insights have direct impact on analysis outcomes. For example, discarding arguments of \texttt{virtualinvoke} may generate false negatives. Suppose \texttt{virtualinvoke} is used to set a key in a \texttt{KeyHolder} instance with some constant: \texttt{virtualinvoke \$r5.<KeyHolder: void setKey(java.lang.String)>("abcd")}. Constant \texttt{abcd} needs to be flagged. 
On the contrary, we observe that arguments of \texttt{virtualinvoke} appearing in assign statements are typically used to describe the state of a variable and can be ignored. In summary, RI-I states that {\em i)} arguments of any \texttt{virtualinvoke} method invocation in an assignment instruction can be regarded as pseudo-influences, and {\em ii)} any constants that influence these arguments can also be discarded.


\subsection{RI-II: Removal of Source Identifiers}

Another type of pseudo-influences due to the clipping of orthogonal method exploration is the identifiers of value sources.
We use an example to illustrate the importance of this insight. For the code below,
a straightforward analysis would flag constant \verb1ENCRYPT_KEY1. However, it is an identifier for retrieving a value from a Java Map data structure, and thus a false positive.

\begin{footnotesize}
\texttt{\$r30 = interfaceinvoke r29.<java.util.Map: java.lang.Object get(java.lang.Object)>("ENCRYPT\_KEY")}
\end{footnotesize}

\noindent
\emph{i) Retrieving values from an external source.}
Static method invocations (\texttt{staticinvoke} in Jimple) in assign statements are typically used to read values from external sources, e.g., Line 15 in Figure~\ref{coding:example}(a):

 \begin{footnotesize}
 \texttt{\$r4 = staticinvoke <Context: java.lang.String getProperty(java.lang.String)>(src)}
 \end{footnotesize}

Variable \texttt{src} refers to the identifier, not the actual value of the key. Thus, it is a pseudo influence. To avoid such pseudo-influences, RI-II discards any arguments of \texttt{staticinvoke} that appear in an assignment. Although \texttt{staticinvoke} may be used to transform a value from one representation to another, it is unlikely to use \texttt{staticinvoke} to transform a constant.

\begin{figure}[hbt]
\small
\centering
\includegraphics[width=0.45\textwidth]{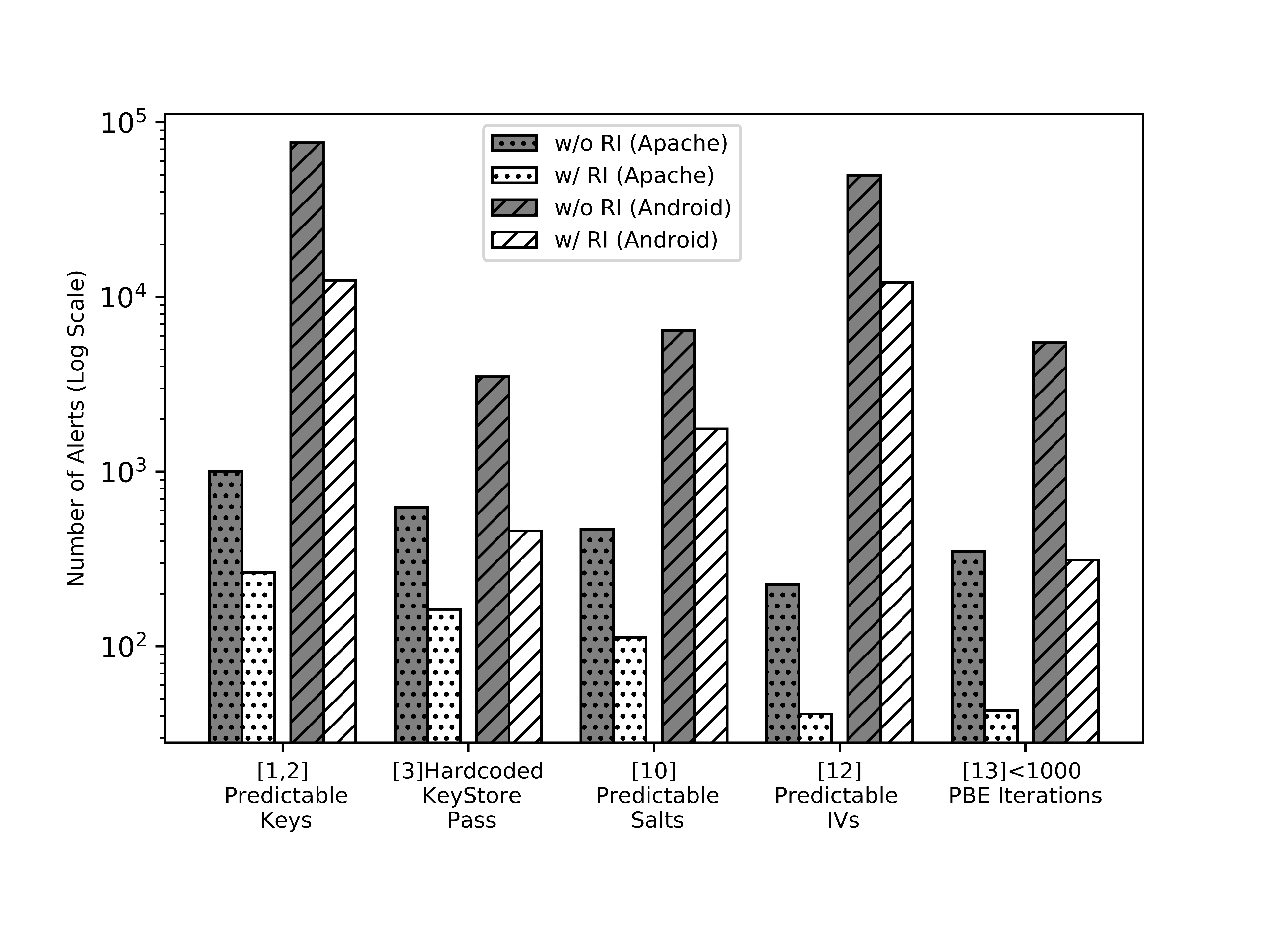}
\vspace{-30pt}
 	\caption{Reduction of false positives with refinement insights in 46 Apache projects (94 root-subprojects) and 6,181 Android apps. Top 6 rules with maximum reductions are shown.}\label{before:after}
 	\vspace{-15pt}
\end{figure}

\begin{figure}[hbt]
\small
\centering
\includegraphics[width=0.42\textwidth]{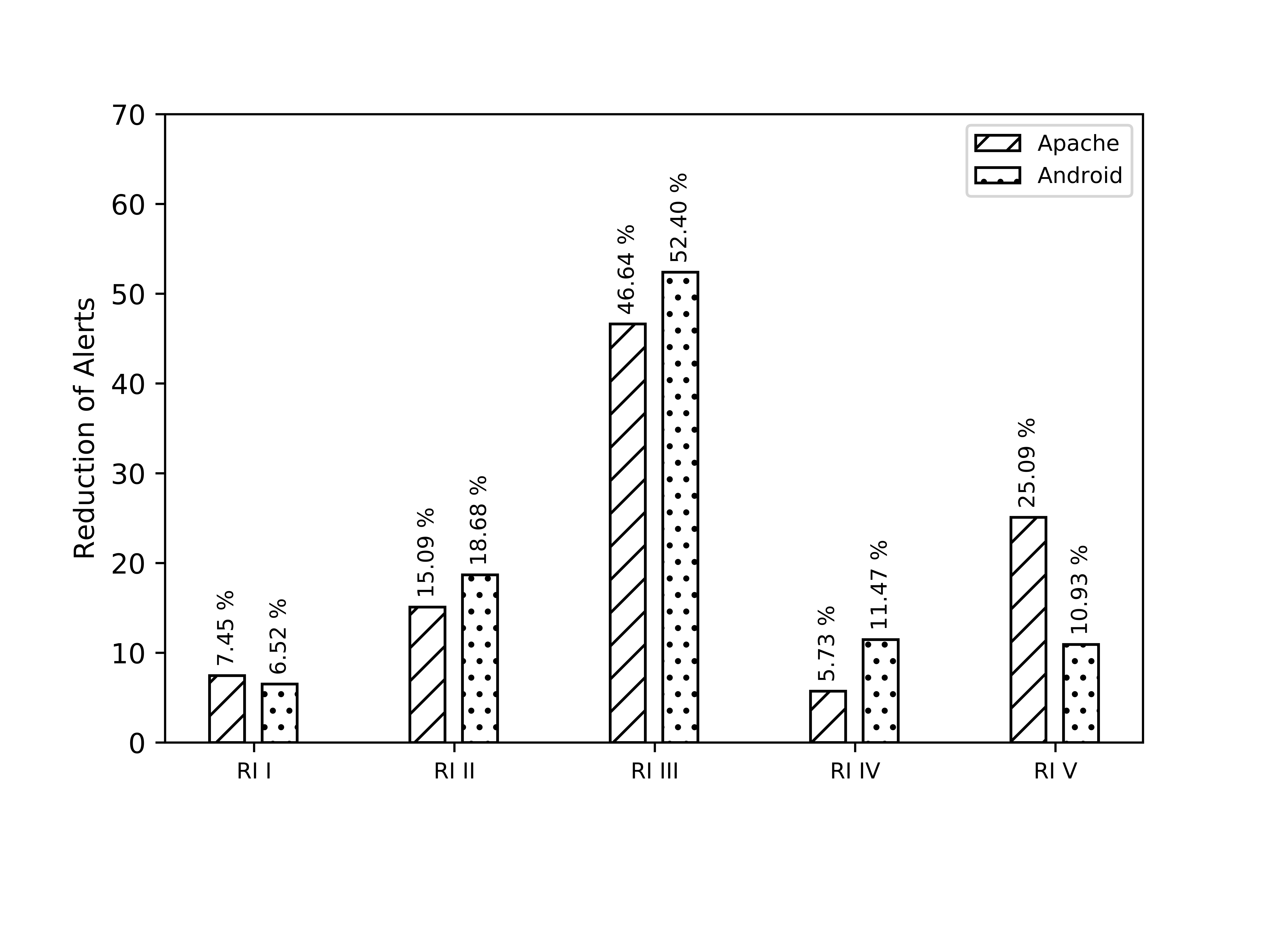}
\vspace{-35pt}
 	\caption{Breakdown of the reduction of false positives due to five of our refinement insights.}\label{breakdown}
 	\vspace{-10pt}
\end{figure}

\subsection{Evaluation of Refinement Methods}\label{analysis-false-alarms}

We compared the numbers of reported alerts before and after employing the five refinement algorithms for 46 Apache projects and 6,181 Android apps. 
Our experiments show that refinement algorithms reduce the total alerts by \apachereduction{} in Apache and \androidreduction{} in Android. For Apache projects, we manually confirmed that all the removed alerts are indeed false positives\footnote{Regarding the validity of the manual analysis, the manual confirmation of alerts was conducted by a second-year Ph.D. student with a prior Master degree in cybersecurity (the second author), under the close guidance of a professor and a senior Ph.D. student (the first author).}. All constant-related rules (including 1, 2, 3, and 12) greatly benefit from the refinements and have significant reduction of irrelevant alerts. Results for top six rules with maximum reductions are shown in Figure~\ref{before:after}. The detailed breakdown is shown in Figure~\ref{breakdown}. The most effective refinement insight for Apache and Android are RI-III (removal of array/collection bookkeeping information).

With refinements enabled, there are a total of \apachealerts{} alerts for the 46 Apache projects. Our careful manual source-code analysis confirms that \apachetruep{} alerts are true positives, resulting in a precision of \precision{}. Out of the 18 false positives, 1 case is due to path insensitivity and 17 to clipping orthogonal explorations (discussed in Section~\ref{sec:discussion}). All experiments reported in the next section were conducted with refinements enabled.

\section{Security Findings and Benchmark Evaluation}
\label{sec:findings}


Our experimental evaluation aims to answer the following questions.

\begin{itemize}[leftmargin=2em]

 \item What are the security findings in Apache Projects? Do Apache projects have too many high-risk vulnerabilities such as hardcoded secrets or MitM vulnerabilities? (Section~\ref{analysis-of-apache})
 \item What are the security findings in Android Apps? Do third-party libraries have high-risk vulnerabilities? (Section~\ref{analysis-of-android})
  \item How does \myname{} compare with CrySL, SpotBugs, and the free trial version of Coverity on benchmarks or real-world projects? (Section~\ref{sec:hestia-fixdroid})
\end{itemize}

\noindent
{\em Selection and pre-processing of programs.}
We selected $46$ popular Apache projects that have crypto API uses. The popularity is measured with the numbers of stars and forks in Github. The maximum, minimum and average Line of Code (LoC) are around $2,571$K (Hadoop), $1.1$K (Commons Crypto) and $402$K, respectively. 
We perform subproject dependency analysis to build DAGs by parsing build scripts. Subproject dependency analysis was automated for \textit{gradle} and \textit{maven}, and was manual for \textit{Ant}. We identified the root-subprojects, which are sub-projects that have no incoming edges on the subproject dependency DAG. We analyzed $94$ root-subprojects in total\footnote{We exclude $15$ test root-subprojects.}.
%
%
We downloaded $6,181$ high popularity Android apps from the Google app market covering $58$ categories. The median value of the number of apps per category is $120$.
We used Soot to decompile .apk files to Java bytecode in order to interface with \myname{}.
We use online APK decompiler~\footnote{http://www.javadecompilers.com/apk} to obtain human-readable source code for manual verification.

\smallskip
\noindent
{\em \myname{} runtime.} We ran 4 concurrent instances of \myname{} in an Intel Xeon(R) X5650 server (2.67GHz CPU and 32GB RAM). Runtime increases with the use of cryptography APIs. For Apache projects, the average runtime was 3.3 minutes with a median of around 1 minute. For Android apps, we terminated unfinished analysis after 10 minutes. The average runtime was 3.2 minutes with a median of 2.85 minutes, including the cutoff ones. 552 (9\%) of 6,181 app's analysis did not finish within 10 minutes, on which \myname{} generated partial results.\footnote{Most of them missed results from Rule 7, which \myname{} runs the last.} 

\begin{table}[]
		\centering
		\captionof{table}{Breakdown of Accuracy in Apache Projects. Duplicates are handled at root-subproject level (total 82 root-subprojects) level. For Rules 1, 2, 3, 8, 10, 12, each constant/predictable value of an array/collection is considered as an individual violation.}
		\label{accuracy:apache}
		\resizebox{\columnwidth}{!}{%
			\begin{tabular}{lccc}
				\Xhline{2\arrayrulewidth}
				\multicolumn{1}{|c|}{\textbf{Rules}}                             & \multicolumn{1}{c|}{\textbf{Total Alerts}} & \multicolumn{1}{c|}{\textbf{\# True Positives}} & \multicolumn{1}{c|}{\textbf{Precision}} \\ \Xhline{2\arrayrulewidth}
				\specialrule{0em}{1pt}{1pt} \Xhline{2\arrayrulewidth}
				
				\multicolumn{1}{|l|}{(1,2) Predictable Keys}              & \multicolumn{1}{c|}{264}              & \multicolumn{1}{c|}{248}                      & \multicolumn{1}{c|}{94.14 \%} \\
				\multicolumn{1}{|l|}{(3) Hardcoded Store Pass}          & \multicolumn{1}{c|}{148}               & \multicolumn{1}{c|}{148}                       & \multicolumn{1}{c|}{100 \%} \\ \Xhline{2\arrayrulewidth}
				\multicolumn{1}{|l|}{(4) Dummy Hostname Verifier}      & \multicolumn{1}{c|}{12}                & \multicolumn{1}{c|}{12}                        & \multicolumn{1}{c|}{100 \%}   \\ 
				\multicolumn{1}{|l|}{(5) Dummy Cert. Validation}       & \multicolumn{1}{c|}{30}               & \multicolumn{1}{c|}{30}                       & \multicolumn{1}{c|}{100 \%}   \\ 
				\multicolumn{1}{|l|}{(6) Used Improper Socket}         & \multicolumn{1}{c|}{4}               & \multicolumn{1}{c|}{4}                       & \multicolumn{1}{c|}{100 \%}   \\ 
				\multicolumn{1}{|l|}{(7) Used HTTP}                     & \multicolumn{1}{c|}{222}               & \multicolumn{1}{c|}{222}                       & \multicolumn{1}{c|}{100 \%}   \\ \Xhline{2\arrayrulewidth}
				\multicolumn{1}{|l|}{(8) Predictable Seeds}             & \multicolumn{1}{c|}{0}         & \multicolumn{1}{c|}{0}                        & \multicolumn{0}{c|}{0\%}        \\
				\multicolumn{1}{|l|}{(9) Untrusted PRNG}                & \multicolumn{1}{c|}{142}               & \multicolumn{1}{c|}{142}                       & \multicolumn{1}{c|}{100 \%}   \\ \Xhline{2\arrayrulewidth}
				\multicolumn{1}{|l|}{(10) Static Salts}            & \multicolumn{1}{c|}{112}               & \multicolumn{1}{c|}{112}                       & \multicolumn{1}{c|}{100 \%}   \\
				\multicolumn{1}{|l|}{(11) ECB mode for Symm. Crypto}       & \multicolumn{1}{c|}{41}               & \multicolumn{1}{c|}{41}                       & \multicolumn{1}{c|}{100 \%}    \\ 
				\multicolumn{1}{|l|}{(12) Static IV}                & \multicolumn{1}{c|}{41}               & \multicolumn{1}{c|}{40}                       & \multicolumn{1}{c|}{97.56 \%} \\ \Xhline{2\arrayrulewidth}
				\multicolumn{1}{|l|}{(13) \textless1000 PBE iterations} & \multicolumn{1}{c|}{43}                & \multicolumn{1}{c|}{42}                        & \multicolumn{1}{c|}{97.67 \%} \\
				\multicolumn{1}{|l|}{(14) Broken Symm. Crypto Algorithm}       & \multicolumn{1}{c|}{86}               & \multicolumn{1}{c|}{86}                       & \multicolumn{1}{c|}{100 \%}    \\ 
				\multicolumn{1}{|l|}{(15) Insecure Asymm. Crypto}    & \multicolumn{1}{c|}{12}                & \multicolumn{1}{c|}{12}                        & \multicolumn{1}{c|}{100 \%}    \\ 
				\multicolumn{1}{|l|}{(16) Broken Hash}                  & \multicolumn{1}{c|}{138}              & \multicolumn{1}{c|}{138}                      & \multicolumn{1}{c|}{100 \%}    \\ \Xhline{2\arrayrulewidth}
				\specialrule{0em}{1pt}{1pt} \Xhline{2\arrayrulewidth}
				\multicolumn{1}{|l|}{\textbf{Total}}                             & \multicolumn{1}{c|}{\textbf{1,295}}              & \multicolumn{1}{c|}{\textbf{1,277}}                      & \multicolumn{1}{c|}{\textbf{98.61 \%}} \\ \Xhline{2\arrayrulewidth}
			\end{tabular}%
		}
\end{table}


\subsection{Security Findings in Apache Projects}\label{analysis-of-apache}

Out of the 46 Apache projects, 39 projects have at least one type of cryptographic misuses and 33 projects have at least two types. Table~\ref{overview:table} summarizes our security findings in screening Apache projects. Predictable keys (Rules 1 and 2), HTTP URL (Rule 7), insecure hash functions (Rule 16), and the insecure PRNGs (Rule 9) are the most common types of vulnerabilities in Apache. As predictable values, we only observed constants for all these rules. We did not observe any predictable seeds under Rule 8. 

\begin{figure}[!h]
\begin{subfigure}[b]{0.45\textwidth}
\lstset{language=Java, 
   basicstyle=\ttfamily\scriptsize,
   keywordstyle=\color{blue}\ttfamily,
   stringstyle=\color{red}\ttfamily,
   commentstyle=\color{cyan}\ttfamily,
   morecomment=[l][\color{magenta}]{\#},
   numbers=left,
   numbersep=2pt,
   tabsize=1,
   escapeinside=||,
   stepnumber=1,
   showstringspaces=false,
   columns=fullflexible,
   xleftmargin=0pt
}
\begin{lstlisting}
 <http:tlsClientParameters |\underline{disableCNCheck="true"}|>
         ...
 </http:tlsClientParameters>
\end{lstlisting}
\vspace{-5pt}
\caption*{(a) A portion of \texttt{https-cfg-client.xml}}
\end{subfigure}
\begin{subfigure}[b]{.45\textwidth}
\lstset{language=Java, 
   basicstyle=\ttfamily\scriptsize,
   keywordstyle=\color{blue}\ttfamily,
   stringstyle=\color{red}\ttfamily,
   commentstyle=\color{cyan}\ttfamily,
   morecomment=[l][\color{magenta}]{\#},
   numbers=left,
   numbersep=2pt,
   tabsize=1,
   escapeinside=||,
   stepnumber=1,
   showstringspaces=false,
   columns=fullflexible,
   xleftmargin=0pt
}
\begin{lstlisting}
...        
} else if (tlsClientParameters.isDisableCNCheck()) {
   verifier = |\underline{new AllowAllHostnameVerifier();}|
}
\end{lstlisting}
\vspace{-5pt}
\caption*{(b) A portion of \texttt{SSLUtils.java}}
\end{subfigure}
\vspace{-5pt}
\caption{Example code in Apache Cxf that disables hostname verification checks by default.}\label{example:rule4}
\end{figure}

\begin{figure}[!h]
\begin{subfigure}[b]{0.45\textwidth}
\lstset{language=Java, 
   basicstyle=\ttfamily\scriptsize,
   keywordstyle=\color{blue}\ttfamily,
   stringstyle=\color{red}\ttfamily,
   commentstyle=\color{cyan}\ttfamily,
   morecomment=[l][\color{magenta}]{\#},
   numbers=left,
   numbersep=2pt,
   tabsize=1,
   escapeinside=||,
   stepnumber=1,
   showstringspaces=false,
   columns=fullflexible,
   xleftmargin=0pt
}
\begin{lstlisting}
 public static String sendUpsRequest(...) {
   ...
   |\underline{http.setAllowUntrusted(true);}|
   ... }
\end{lstlisting}
\caption*{(a) A portion of \texttt{UpsServices.java}}
\end{subfigure}
\begin{subfigure}[b]{.45\textwidth}
\lstset{language=Java, 
   basicstyle=\ttfamily\scriptsize,
   keywordstyle=\color{blue}\ttfamily,
   stringstyle=\color{red}\ttfamily,
   commentstyle=\color{cyan}\ttfamily,
   morecomment=[l][\color{magenta}]{\#},
   numbers=left,
   numbersep=2pt,
   tabsize=1,
   escapeinside=||,
   stepnumber=1,
   showstringspaces=false,
   columns=fullflexible,
   xleftmargin=0pt
}
\begin{lstlisting}
 SSLContext getSSLContext(String alias, boolean trustAny) {
    ...  
    TrustManager[] tm;
    if (trustAny) {
      tm = |\underline{SSLUtil.getTrustAnyManagers()}|;  } ... }
   
\end{lstlisting}
\caption*{(b) A portion of \texttt{SSLUtil.java}}
\end{subfigure}
\caption{Example code in Apache Ofbiz that enables trusting all certificates by default while invoking UPS service.}\label{example:rule5}
\end{figure}

\subsubsection{Vulnerabilities from Predictable Secrets}

16 Apache projects (37 sub-rootprojects) have hardcoded keys (Rule 1, 2). Three (Meecrowave, Kylin, and Cloudstack) of them use hardcoded symmetric keys (Rule 1). Meecrowave uses \textit{DESede} (i.e., Triple DES\footnote{Triple DES itself is considered insecure. OpenSSL removed the support of Triple DES. NIST recommended moving to AES as soon as possible~\cite{NIST-Triple-DES-2017}.}) for obfuscation purpose. Unfortunately, deterministic keys make it trivial to break the obfuscation. Kylin (635 Forks, 1325 Stars) uses \textit{AES} to encrypt user passwords. However, using hardcoded keys make these passwords vulnerable. In Apache Cloudstack, it appears that hardcoded keys are used in test code, which is accidentally packaged with the production code. 

For Rule 2, we found that most of the hardcoded passwords in PBE serve as the default. The most common default password for PBE is \texttt{masterpassphrase} (e.g., Ambari and Knox). Manifoldcf uses \texttt{NowIsTheTime}. 
Setting PBE code to take the default hardcoded passwords without sufficient warnings are risky. Distributions using the default configuration are susceptible to the recovery of the plaintext password by an attacker who has the access to the PBE ciphertext. 
Apache Ranger (165 forks, 155 stars) uses a hardcoded password as default for PBE for all distributions. Its installation Wiki does not mention anything about it. System administrators unaware of this setup are likely not to change the default.
This coding practice significantly weakens the security guarantee of PBE.

For Rule 3, most common hardcoded passwords for KeyStores (for storing private keys) are \texttt{changeit} (e.g., Tomcat, Knox, Judi, Ofbiz and Wss4j) and \texttt{none} (e.g., Knox, Hive and Hadoop). Most of them are set as default. There are 9 projects that have both predictable keys (Rule 1, 2) and hardcoded KeyStore passwords (Rule 3), indicating persistent insecure coding styles.

\noindent
{\bf Insecure common practices.}
During manual analysis, we found three types of insecure common practices in Apache projects for storing secrets: {\em i)} hard-coding default keys or passwords in the source code, {\em ii)} storing plaintext keys or passwords in configuration files, {\em iii)} storing encrypted passwords in configuration files with decryption keys in plaintext in source code or configuration.
Java provides a special security APIs (e.g., \verb1Callback1 and \verb1CallbackHandler1) to prompt users for secrets (e.g., passwords). However, none of these projects provides any code to support this option.

Sysadmins are forced to store plaintext passwords in the filesystem unless they personally modify the code. The biggest danger that these insecure secret-storage practices bring to users is probably the inflated sense of security and not being able to see the actual risks. 



\subsubsection{Vulnerabilities from SSL/TLS MitM}

Man-in-the-Middle (MitM) vulnerabilities are high risk in our threat model. 5 Apache projects (8 root-subprojects) have dummy hostname verifiers that accept any hostnames (Rule 4), including Spark (15086 forks, 16324 stars), Ambari (814 forks, 778 stars), Cxf (706 forks, 398 stars), Ofbiz, and Meecrowave. 6 Apache projects have dummy trust managers that trust any certificates (Rule 5), including Spark, Ambari, Cloudstack, Qpid-broker, Jclouds, and Ofbiz. It appears that most projects offer them as an additional connectivity option.

Our manual analysis reveals that some projects set this insecure implementation as default (e.g., Figure~\ref{example:rule4} and Figure~\ref{example:rule5}). In Figure~\ref{example:rule5}, we see that Ofbiz uses insecure SSL/TLS configurations by default while using UPS (a shipping company) service.
When plain sockets are used, it is recommended to verify the hostname manually. We found 3 projects that do not follow this rule and accept any arbitrary hostnames. We also found 7 projects (24 root-subprojects) that occasionally use the HTTP protocol for communication.


\lstset{language=Java, 
   basicstyle=\ttfamily\scriptsize,
   keywordstyle=\color{blue}\ttfamily,
   stringstyle=\color{red}\ttfamily,
   commentstyle=\color{cyan}\ttfamily,
   morecomment=[l][\color{magenta}]{\#},
   numbers=left,
   numbersep=2pt,
   escapeinside=||,
   tabsize=1,
   stepnumber=1,
   showstringspaces=false,
   columns=fullflexible,
   xleftmargin=0pt
}

\begin{lstlisting}[caption={A vulnerable code snippet from Apache Ranger to demonstrate various security issues},label={code:ranger}]
PBEKeySpec getPBEParameterSpec(String password) throws Throwable {
    MessageDigest md = MessageDigest.getInstance(|\underline{MD\_ALGO}|); // MD5
    byte[] saltGen = |\underline{md.digest(password.getBytes())}|;
    byte[] salt = new byte[SALT_SIZE];
    System.arraycopy(|\underline{saltGen}|, 0, |\underline{salt}|, 0, SALT_SIZE);
    int iteration = |\underline{password.toCharArray().length + 1}|;
    return new PBEKeySpec(password.toCharArray(), salt, iteration);
}
\end{lstlisting}

\subsubsection{Medium and Low Severity Vulnerabilities}

It is important to be aware of the medium and low-risk vulnerabilities in the system and to recognize that the risk levels may increase under different adversarial models.

We found hardcoded salts in 4 projects including Apache Ranger, Manifoldcf, Juddi, and Wicket. We also observe the use of ECB mode in AES in 5 projects and predictable IVs in 2 projects with a total of 40 occurrences.
We found 5 projects that use PBE with less than 1,000 iterations (Rule 13). Ranger and Wicket projects use 17 iterations for PBE; and Incubator-Taverna-Workbench and Juddi projects use 20 iterations, much fewer than the required 1,000. 

Listing~\ref{code:ranger} shows a code snippet from Ranger, which has multiple issues. The number of iterations is proportional to the password size (Line 6), which is far less than $1,000$. In addition, this code offers a timing side-channel. An adversary capable of measuring PBE execution time (e.g., in multi-tenant environments) may learn the length of the password. This information can substantially decrease the difficulty of dictionary attacks. Another issue is that the salt is computed as the MD5 hash of the password (Lines 2-3). An adversary obtaining the salt may quickly recover the password. The salt's dependence on the password itself also breaks the indistinguishability requirement of PBE under chosen plaintext attack.  

We found various occurrences of Blowfish, DES, and RC4 ciphers for Rule 14. Under Rule 15, we found 3 occurrences of using default key size of 1024 and 9 other occurrences that explicitly initialize the key size to 1024. 
$23$ projects use \texttt{java.util.Random} as a PRNG (Rule 9), where two of them set static seeds to \texttt{java.util.Random}. 
We do not observe any deterministic seed to a \texttt{java.security.SecureRandom} (Rule 8). 


\lstset{language=Java, 
   basicstyle=\ttfamily\scriptsize,
   keywordstyle=\color{blue}\ttfamily,
   stringstyle=\color{red}\ttfamily,
   commentstyle=\color{cyan}\ttfamily,
   morecomment=[l][\color{magenta}]{\#},
   numbers=left,
   numbersep=2pt,
   tabsize=1,
   stepnumber=1,
   escapeinside=++,
   showstringspaces=false,
   columns=fullflexible
}
\begin{lstlisting}[caption={An example of only checking the expiration (\texttt{checkValidity}) of self-signed certificates in \textit{Yahoo Finance (TWStock)} Android app. The base package name (\texttt{com.softmobile}) of this class indicates that the vulnerable code comes from a third-party library.},label={yahoo:finance}]
void checkServerTrusted(X509Certificate[] chain, String str){
  if (chain == null || chain.length != 1) {
      this.f7654a.checkServerTrusted(chain, str);
  } else {
     //Lack of signature verification and others
     +\underline{chain[0].checkValidity()}+;}}
\end{lstlisting}

\lstset{language=Java, 
   basicstyle=\ttfamily\scriptsize,
   keywordstyle=\color{blue}\ttfamily,
   stringstyle=\color{red}\ttfamily,
   commentstyle=\color{cyan}\ttfamily,
   morecomment=[l][\color{magenta}]{\#},
   numbers=left,
   numbersep=2pt,
   escapeinside=||,
   tabsize=1,
   stepnumber=1,
   showstringspaces=false,
   columns=fullflexible
}
\begin{lstlisting}[caption={An example of ignoring exceptions in \texttt{checkServerTrusted} in \textit{Sina Finance} Android app.},label={sina:finance}]
void checkServerTrusted(X509Certificate[] chain, String str){
   try {
     this.f7427a.checkServerTrusted(chain, str);
   } catch (CertificateException e) {}} //Ignores exception
\end{lstlisting}

\lstset{language=Java, 
   basicstyle=\ttfamily\scriptsize,
   keywordstyle=\color{blue}\ttfamily,
   stringstyle=\color{red}\ttfamily,
   commentstyle=\color{cyan}\ttfamily,
   morecomment=[l][\color{magenta}]{\#},
   numbers=left,
   numbersep=2pt,
      escapeinside=||,
   tabsize=1,
   stepnumber=1,
   showstringspaces=false,
   columns=fullflexible
}
\begin{lstlisting}[caption={The use of \texttt{SSLSocket} without manual hostname verification in \textit{ProTaxi Driver} Android app.},label={socket:hostname}]
try {
  SSLContext instance = SSLContext.getInstance("TLS");
  ...
  this.webSocketClient
      .setSocket(instance.getSocketFactory().|\underline{createSocket()}|);
} catch (Throwable e) { ... }
this.webSocketClient.|\underline{connect()}|;
\end{lstlisting}

\begin{scriptsize}
\begin{table*}[!hbt]

\setlength{\tabcolsep}{0.5em}

\footnotesize
\caption{Experimental results on the \bench{} basic and \bench{} advanced benchmarks with CrySL, Coverity, SpotBugs and \myname{}. GTP stands for the ground truth positives. TP, FP, and FN are the number of true positives, false positives, false negatives in a tool's output, respectively. Pre. and Rec. represent precision and recall, respectively. Tools are evaluated on 6 common rules (out of our 16 rules), i.e., the maximum common subset of all tools. For these 6 rules, there are 6 correct cases (i.e., true negatives) in basic and 3 correct cases in advanced, which are used for computing FPRs. Total alerts $=$ TP $+$ FP. }
 \label{tab:benchmark}
\begin{tabular}{|c|ccccccc|ccc|ccc|ccc|ccc|cccc|}
\hline
{\textbf{Tools}} & \multicolumn{7}{c|}{\textbf{\bench{}: Basic}} & \multicolumn{16}{c|}{\textbf{\bench{}: Advanced}} \\ \cline{2-24} 
                               & \multicolumn{3}{c|}{\textbf{GTP:14}}                                               & \multicolumn{4}{c|}{\textbf{Summary}}                                                                                  & \multicolumn{3}{c|}{\begin{tabular}[c]{@{}c@{}}\textbf{Inter-Proce.} \\ \textbf{(Two)} \\ \textbf{GTP: 13}\end{tabular}} & \multicolumn{3}{c|}{\begin{tabular}[c]{@{}c@{}}\textbf{Inter-Proce.}\\ \textbf{(Multiple)}\\ \textbf{GTP: 13}\end{tabular}} & \multicolumn{3}{c|}{\begin{tabular}[c]{@{}c@{}}\textbf{Field} \\ \textbf{Sensitive}\\ \textbf{GTP: 13}\end{tabular}} & \multicolumn{3}{c|}{\begin{tabular}[c]{@{}c@{}}\textbf{FP Test/}\\ \textbf{Correct Uses}\\ \textbf{GTP: 3}\end{tabular}} & \multicolumn{4}{c|}{\textbf{Summary}}                                                                                  \\ \cline{2-24} 
                               & \textbf{TP}                     & \textbf{FP}                    & \multicolumn{1}{c|}{\textbf{FN}} & \textbf{FPR}                      & \textbf{FNR}                      & \textbf{Pre.}                      & \textbf{Rec.}                       & \textbf{TP}                           & \textbf{FP}                          & \textbf{FN}                            & \textbf{TP}                            & \textbf{FP}                           & \textbf{FN}                             & \textbf{TP}                          & \textbf{FP}                          & \textbf{FN}                           & \textbf{TP}                         & \textbf{FP}                         & \textbf{FN}                          & \textbf{FPR}                      & \textbf{FNR}                       & \textbf{Pre.}                      & \textbf{Rec.}                       \\ \hline
CrySL\cite{DBLP:conf/ecoop/KrugerS0BM18}                          & {10}                     & {4}                     & \multicolumn{1}{c|}{4}  & {40.00}                    & {28.57}                    & {{\bf 71.43}}                     & {{\bf 71.43}}                      & {10}                           & {3}                          & {3}                             & {0}                             & {12}                           & {13}                             & {0}                           & {1}                           & {13}                           & {0}                          & {2}                          & {3}                           & {85.71}                    & {76.19}                     & {{\bf 35.71}}                     & {{\bf 23.81}}                      \\
{Coverity\cite{coverity}} & {13} & {0} & \multicolumn{1}{c|}{1}  & {0.00} & {7.14} & {{\bf 100.0}} & {{\bf 92.86}} & {3}  & {0}       & {10}       & {3}         & {0}        & {10}        & {1}       & {0}       & {12}      & {0}      & {0}      & {3}      & {0.00} & {83.33} & {{\bf 100.0}} & {{\bf 16.67}} \\
SpotBugs\cite{spotbugs}                       & {13}                     & {0}                     & \multicolumn{1}{c|}{1}  & {0.00}                     & {7.14}                     & {{\bf 100.0}}                    & {{\bf 92.86}}                      & {0}                            & {0}                           & {13}                            & {3}                             & {10}                           & {10}                             & {0}                           & {0}                           & {13}                           & {0}                          & {0}                          & {3}                           & {76.92}                    & {92.86}                     & {{\bf 23.08}}                     & {{\bf 7.14}}                       \\
\myname{}                        & {14}                     & {0}                     & \multicolumn{1}{c|}{0}  & {0.00}                     & {0.00}                     & {{\bf 100.0}}                    & {{\bf 100.0}}                     & {13}                           & {0}                           & {0}                             & {13}                            & {0}                            & {0}                              & {13}                          & {0}                           & {0}                            & {3}                          & {0}                          & {0}                           & {0.00}                     & {0.00}                      & {{\bf 100.0}}                    & {{\bf 100.0}}                     \\ \hline
\end{tabular}
\end{table*}
\end{scriptsize}

\subsection{Security Findings in Android Apps}\label{analysis-of-android}

\noindent
{\em Violations in apps or in libraries?}  We distinguished app's own code from libraries by using the package information from \texttt{AndroidManifest.xml}.\footnote{An .apk contains both the app code and the libraries.} Android also uses it during \texttt{R.java} file generation (robust against obfuscation). We found that on average \textbf{95\% of the detected vulnerabilities come from libraries} (Table~\ref{table:app-violation-source-percent}). This result extends the observation from 7 types of vulnerabilities (reported in~\cite{DBLP:conf/ccs/0001BD16}) to 16.

Table~\ref{table:app-violation-source-percent} shows the distribution of vulnerability sources for each rule.
For hardcoded KeyStore passwords (Rule 3), all violations come from libraries. Most frequent hardcoded KeyStore password is \texttt{notasecret}, which is used to access certificates and keys in Google libraries (e.g., \texttt{*.googleapis.GoogleUtils},  \texttt{*.googleapis.*.GoogleCredential}).

\begin{table}[!htbp]
\caption{Distribution of vulnerabilities in Android apps.}
\label{table:app-violation-source-percent}
\centering
\scriptsize
\begin{tabular}{|l|l|l|l|l|}
\hline
              & \textbf{Library} & \textbf{Library} & \textbf{App Itself} & \textbf{Total} \\
              & \textbf{(Total)} & \textbf{(Unique)} & & \\       
              \hline \hline
(1,2) Predictable Keys    & 11,634 (93.4\%) & 5,940 & 823 (6.6\%) & 12,457\\
(3) Hardcoded Store Password & 431 (94.1\%) & 170 & 27 (5.8\%) & 458 \\ \hline
(4) Dummy Hostname Verifier & 1,148 (99.3\%) & 51 & 7 (0.7\%) & 1,155 \\
(5) Dummy Cert. Validation & 3,715 (96.3\%) & 1,317 & 141 (3.7\%) & 3,856\\ 
(6) Used Improper Socket & 270 (99.6.4\%) & 13 & 1 (0.4\%) & 271\\ 
(7) Used HTTP            & 7,687 (92.5\%) & 2,105 & 623 (7.5\%) & 8,321 \\ \hline 
(8) Predictable Seeds & 522 (96.0\%) & 101 & 22 (4.0\%) & 544\\ 
(9) Untrusted PRNG & 26,312 (91.7\%) & 8,679 & 2,393 (8.3\%) & 36,223 \\ \hline
(10) Predictable Salts  & 1,638 (93.2\%) & 774 & 119 (6.8\%) & 1,757 \\ 
(11) ECB in Symm. Crypto  & 1657 (93.1\%) & 682 & 123 (6.9\%)  & 1,780 \\ 
(12) Predictable IVs    & 11,357 (94.2\%) & 6,048 & 692 (5.8\%) & 12,089 \\ \hline
(13) \textless 1000 PBE iterations & 294 (94.2\%) & 129 & 18 (57.8\%) & 312 \\ 
(14) Broken Symm. Crypto  & 1,668 (95.8\%) & 753 & 74 (4.2\%) & 1,742 \\ 
(15) Insecure Asymm. Crypto  & 4 (3.6\%) & 3 & 107 (96.4\%) & 111 \\
(16) Broken Hash & 49,257 (99.0\%) & 7509 & 496 (1.0\%) & 49,769 \\ \hline 
\textbf{Total} & \textbf{117,594 (95.40\%)} & \textbf{34,274} & \textbf{5,666 (4.60\%)} & 130,845 \\ \hline 
\end{tabular}
\end{table}


Besides Google, other high-profile library sources include Facebook, Apache, Umeng, and Tencent (Table~\ref{table:violation-by-packagename}). These libraries frequently appear in different applications. We distinguished these libraries using base packages and ignored obfuscations.

\begin{table}[t]
\caption{Violations in 5 popular libraries (manually confirmed).}
\label{table:violation-by-packagename}
\centering
\footnotesize
\begin{tabular}{|l|l|}
\hline 
\textbf{Package name }                                        & \textbf{Violated rules}                                                                                                                \\ \hline \hline 
com.google.api  & 3, 4, 5, 7 \\ \hline
com.umeng.analytics  & 7, 9, 12, 16 \\ \hline
com.facebook.ads   & 5, 9, 16\\ \hline
org.apache.commons & 5, 9, 16 \\ \hline
com.tencent.open & 2, 7, 9 \\ \hline
\end{tabular}
\vspace{-10pt}
\end{table}

\noindent
{\em Overview of other Android findings.}
We found exposed secrets, similar to Apache projects. Table~\ref{overview:table} summarizes the discovered vulnerabilities in Android applications. The categories of untrusted PRNG (Rule 9) and broken hash (Rule 16) have the most violations. Interestingly, we observed 544 cases of predictable seeds (Rule 8). 13 cases of them used time-stamps from \texttt{<java.lang.System: long currentTimeMillis()>} API calls. 

{\bf Compared with Apache projects, Android apps have higher percentages of SSL/TLS API misuses (Rules 4, 5 and 6) and HTTP use (Rule 7).} For example, 25.30\% of Android apps have dummy trust manager (Rule 5), which is more than 2 times of the number in Apache (11.70\%) as shown in Table~\ref{overview:table} in Appendix. 

Our analysis can detect sophisticated cases that Google Play's built-in screening is likely to miss. We give code snippets for such cases (Listing~\ref{yahoo:finance},~\ref{sina:finance},~\ref{socket:hostname}). \myname{} detects a case where developers allow unpinned self-signed certificates with a mere expiration check, as shown in Listing~\ref{yahoo:finance}.
Another case is where developers ignore the exception in \texttt{checkServerTrusted} method as shown in Listing~\ref{sina:finance}. 
In addition, \myname{} detects 271 occurrences of improper use of \texttt{SSLSocket} without manual Hostname verification in 210 apps. One such example is shown in  Listing~\ref{socket:hostname}, where \texttt{SSLSocket} is used in \texttt{WebSocketClient} without manually verifying the hostname~\footnote{Guide for the correct use can be found at \url{https://developer.android.com/training/articles/security-ssl\#WarningsSslSocket}.}. In comparison, Google Play's inspection appears to only detect obvious misuses~\cite{google-auto-checking-bypass}.

Grouping security violations by app popularity or category did not show substantial differences across groups.





\subsection{Comparison with Existing Tools}\label{sec:hestia-fixdroid}

We compare the accuracy and runtime of \myname{} with three existing tools, i.e., CrySL~\cite{DBLP:conf/ecoop/KrugerS0BM18},  Coverity~\cite{coverity}, and SpotBugs~\cite{spotbugs}\footnote{CryptoLint's code is unavailable.}.  During our experiments, we use CrySL 1.0 (commit id \textit{10e86fdb}), SpotBugs 3.0.1 (from SWAMP) and the results from Coverity was obtained before Jan 07, 2019.

\noindent
{\bf Benchmark preparation.} First, we\footnote{The person (third author) who led the benchmark design is different from the person (first author) who implemented \myname{}.} had to construct \bench{}, a comprehensive benchmark for comparing the quality of cryptographic vulnerability detection tools. Regarding the existing benchmark DroidBench~\cite{DBLP:conf/pldi/ArztRFBBKTOM14}, \textit{i)} DroidBench does not cover cryptographic APIs, \textit{ii)} the free web version of Coverity requires source code, however DroidBench only contains APK binaries.

\bench{} covers all 16 cryptographic rules specified in Table~\ref{table:rules}. There are 38 basic test cases and 74 advanced test cases. The basic benchmark contains 25 straightforward API misuses and 13 correct API uses (i.e., true negative cases). The advanced cases have more complex scenarios, including 42 inter-procedural cases\footnote{21 cases involve two methods and 21 cases involve more than two methods.}, 20 field-sensitive cases, 9 false positive test cases (for evaluating the ability of recognizing irrelevant elements), and 3 correct API uses (i.e., true negative cases). Figures~\ref{rules:testcases} and~\ref{apis:testcases}  in Appendix show the distributions of test cases per rule and per API, respectively. Augmenting the benchmark with more test cases is our ongoing work. See Github for the most updated version~\url{https://github.com/CryptoGuardOSS/cryptoapi-bench}.



\noindent
{\bf Benchmark comparison.} 
To maintain fairness in our comparison, we only report the benchmark results for the six shared rules (1, 2, 3, 11, 14, 16) that are covered by all the tools,  CrySL~\cite{DBLP:conf/ecoop/KrugerS0BM18},  Coverity~\cite{coverity},  SpotBugs~\cite{spotbugs}, and ours. Due to the lack of documentation, we had to infer a tool's coverage based on whether or not it ever generates any alert in that category. We show the results in Table~\ref{tab:benchmark}. 

For the basic benchmark, SpotBugs and Coverity perform well, but not 
CrySL. Our investigation reveals that CrySL's false positives are mainly due to their rules being overly strict. For example, CrySL would raise an alert if the password for PBE is derived from a \verb1String1 typed variable, or a symmetric key is not generated by a key generator. It cannot recognize 4 correct API uses in the evaluation (out of 9). The root cause for this overly specific definitions of security is likely the CrySL's language restrictions on constraint definitions. 
For the advance benchmark, both CrySL and SpotBugs generate false positives, when a variable is passed through multiple methods. For all cases, Coverity has zero false positives, likely because of the use of symbolic execution and/or path-sensitive analysis\footnote{Coverity is close sourced, so we are unable to confirm.}. However, Coverity misses multiple advanced vulnerability scenarios (for rules that it does cover in the basic benchmark). 


Table~\ref{table:totalbench} in Appendix presents the comparison for all 16 rules (not just the 6 common rules as in above). When testing all 16 rules, \myname{} failed to report 11 misuses (i.e., false negatives). We discuss the causes in Section~\ref{sec:discussion}. 



\noindent
{\bf Runtime comparison.} We ran CrySL and \myname{} on 10 randomly selected Apache root-subprojects. Unfortunately, CrySL crashed and exit prematurely for 7 of them. For the 3 completed projects, CrySL is slower, but comparable on 2 projects (5 vs. 3 seconds, 25 vs. 19 seconds). However, it is 3 orders of magnitude slower than \myname{} on \verb1kerbaros-codec1.\footnote{Reported runtime is the average of three runs.}

 We choose not to compare with SpotBugs -- the comparison would not be meaningful, because its analysis is mostly based on the syntactical matching of source code to known bug patterns~\cite{DBLP:journals/sigplan/HovemeyerP04, DBLP:conf/issre/RutarAF04}. For the free web version of Coverity, we are unable to obtain its runtime.

\noindent
{\em Summary of experimental findings.}

\begin{itemize}[leftmargin=2em]

 \item 

Our refinement algorithms are effective. They bring an \apachereduction{} reduction in alarms for Apache projects and an \androidreduction{} reduction for Android applications. We manually confirmed that all the removed alerts are indeed false positives.
Manually examining the remaining \apachealerts{} Apache alerts (after refinements) confirms our precision of \precision{}.

 \item 

39 out of the 46 Apache projects have at least one type of cryptographic misuses and 33 have at least two types. There is a widespread insecure practice of storing plaintext passwords in code or in configuration files. Insecure uses of SSL/TLS APIs are set as the default configuration in some cases. 
  
 \item 

5,596 (91\%) out of the 6,181 Android apps have at least one type of cryptographic misuses and 4,884 (79\%) apps have at least two types. 95\% of the vulnerabilities come from the libraries that are packaged with the applications. Some libraries are from large software firms. 
\myname{}'s detection for SSL/TLS API misuses is more comprehensive than the built-in screening offered by Google Play.

\item
In terms of detecting complex misuses, \myname{} outperforms all leading solutions (in Table~\ref{tab:benchmark}). It substantially outperforms CrySL in terms of robustness and runtime.

\end{itemize}

\section{Discussion}
\label{sec:discussion}

\noindent
{\bf Code correction.}
Most of the Apache developers' responses to our vulnerability disclosure reports were prompt and insightful. We highlight the feedback from some projects.
Apache Spark promised to remove the support of dummy hostname verifier and trust store.
Ofbiz promised to fix the reported issues of constant IVs and KeyStore passwords. Apache Ranger already fixed our report of constant default values for PBE~\cite{ranger-issue-key} and insecure cryptographic primitives~\cite{ranger-issue-crypto}. Regarding MD5, Apache Hadoop justifies that its MD5 use is for the per-block checksums for Hadoop file systems (HDFS)'s consistency and the setup does not assume the presence of active adversaries.  

For some cases, developers explained that certain operational constraints (e.g., backward compatibility for clients) prevent them from fixing the problems. 
For example, Apache Tomcat server has to use MD5 in its digest authentication code, because major browsers do not support secure hash functions (as defined in RFC 7616). However, digest authentication is rarely used in the wild\footnote{https://security.stackexchange.com/questions/152935/why-is-there-no-adoption-of-rfc-7616-http-digest-auth}.

The thorniest issue is secret storage. One justification for developers' choice of storing plaintext passwords or keys in file systems is for supporting humanless environments (e.g., automated scripts to manage services). However, first, not all deployment scenarios are server farms in a humanless environment. Projects should also provide the secure option, which is to use Java callback to prompt human operators for passwords which can be used to unlock/generate other passwords or keys on the fly. Second, not properly disclosing and documenting the insecure configurations does a great disservice to the project's users. 



\noindent
{\bf Our limitations.} 
No static analysis tool is perfect. \myname{} is no exception. We discuss the detection limitations of \myname{} and future improvements.

\noindent
{\em False positives.} One source of false positives comes from the path insensitivity. For example, \myname{} raises an alert if the variable \texttt{iteration} is assigned with a value of $0$ for the following code snippet (from project \textit{jackrabbit-oak}). However, this alert is a false positive, since this assignment is on an infeasible path.

\begin{footnotesize}
\begin{verbatim}
int iteration = 0;
...
if (iteration < NO_ITERATION) { // NO_ITERATION = 1
    iteration = DEFAULT_ITERATION;
}
\end{verbatim}
\end{footnotesize}
Another source of false positives is the clipping of orthogonal exploration, which can be substantially reduced by re-configuring the prototype to explore deeper. However, this will impact the runtime.
In addition, \myname{} detects the existence of API misuses in a code base but does not verify that the vulnerable code will be triggered at runtime. This issue is a general limitation of static program analysis. 
Apache Spark confirmed insecure PRNG uses, but stated that the affected code regions are not security critical.\footnote{It is unclear why Spark chose to use insecure PRNG, even for non-security purposes.} However, eliminating this type of alerts is difficult, if possible at all, as the analysis needs to be aware of custom defined security criteria (e.g., what constitutes critical security) with in-depth knowledge about project semantics.

\noindent
{\em False negatives.} For the full benchmark evaluation in Table~\ref{table:totalbench} in Appendix, \myname{} has 11 false negatives (i.e., missed detection). All these cases are due to our refinements after clipping orthogonal explorations.
For example, RI-II would ignore \texttt{6A5B7C8A} as a pseudo-influence from the following instruction, if orthogonal explorations are clipped to explore \textit{parseHexBinary} method. 
\begin{footnotesize}
 \texttt{byte[] key = DatatypeConverter.parseHexBinary("6A5B7C8A").}
\end{footnotesize}

These false negatives can be avoided by increasing the depth of the orthogonal exploration.
However, these conversions are mostly required to absorb values from external sources (e.g., file system, network). Any such conversions of static values under the rules of Table~\ref{table:rules} are highly unlikely. Outside the benchmark, we did not observe any such cases during our manual investigation of Apache alerts. 
We also manually investigated 5 randomly selected Apache projects for false negatives from within orthogonal invocations (due to clipping) and did not observe any.

\myname{} runs the intra-procedural forward slicing for Rules 6 and 15, where an inter-procedural forward slicing could potentially improve the coverage.
For Rule 15, this change might not make much difference, as \texttt{KeyPairGenerator} creation and its initialization usually occur in the same method.
For Rule 6, our current implementation ignores the direct sub-classes of \texttt{SSLSocketFactory} to avoid false positives. Inter-procedural slicing could extend the analysis to the sub-classes.

Another limitation of our work is that the coverage of our benchmark needs further improvement, in terms of incorporating path sensitive test cases and increasing the diversity of APIs involved. Our ongoing work includes enhancing \bench{}.






\section{Related Work}

\noindent
{\em Tools to detect cryptographic misuses.} Cryptographic misuse detection tools are typically constructed into two broad groups, i.e., static analysis (e.g., CryptoLint~\cite{DBLP:conf/ccs/EgeleBFK13}, MalloDroid~\cite{DBLP:conf/ccs/FahlHMSBF12}, FixDroid~\cite{DBLP:conf/ccs/NguyenWABWF17}, CogniCrypt~\cite{DBLP:conf/kbse/KrugerNRAMBGGWD17} and CrySL~\cite{DBLP:conf/ecoop/KrugerS0BM18}) and dynamic analysis (e.g., SMV-Hunter~\cite{DBLP:conf/ndss/SounthirarajSGLK14}, AndroSSL~\cite{DBLP:conf/fps/GagnonFFDOB15} and K-Hunt~\cite{DBLP:conf/ccs/LiLCZG18}). For example, MalloDroid~\cite{DBLP:conf/ccs/FahlHMSBF12} uses a list of known insecure implementations of \texttt{HostnameVerifier} and \texttt{TrustManager} to screen Android apps. 
In~\cite{DBLP:conf/icse/JohnsonSMB13}, authors showed that generating false positives is one of the most significant barrier to adopt static analysis tools. This false positive problem also exists in anomaly and intrusion detection systems~\cite{DBLP:journals/cn/LippmannC00,Yao-book-2017}. When screening large projects, virtually all static slicing solutions in this space (e.g.,~\cite{DBLP:conf/ccs/EgeleBFK13}) might generate a non-negligible amount of false positives. Contextual refinements similar to \myname{}'s is necessary to achieve high precision in practice. In terms of the coverage, \myname{} covers more rules than CryptoLint~\cite{DBLP:conf/ccs/EgeleBFK13}, CrySL~\cite{DBLP:conf/ecoop/KrugerS0BM18} and MalloDroid~\cite{DBLP:conf/ccs/FahlHMSBF12} combined.

Other misuse detection tools (e.g., FixDroid~\cite{DBLP:conf/ccs/NguyenWABWF17} and CogniCrypt~\cite{DBLP:conf/kbse/KrugerNRAMBGGWD17}) were mainly built for the user-experience study with the goal of making detection tools developer-friendly, as opposed to a deployment-quality screening solution. For example, FixDroid focuses on providing real-time feedback to developers. CogniCrypt's~\cite{DBLP:conf/kbse/KrugerNRAMBGGWD17} focus is on code generation (in Eclipse IDE) for several common cryptographic tasks (e.g., data encryption).


Dynamic analysis tools are complementary to static analysis ones. Most of them use a simple static analysis to first narrow-down the number of potential apps for dynamic analysis. For example, SMV-Hunter~\cite{DBLP:conf/ndss/SounthirarajSGLK14} looks for apps that contain any custom implementation of \texttt{X509TrustManager} or \texttt{HostNameVerifier} for initial screening.

\noindent
{\em Other static analysis tools.}  TaintCrypt~\cite{DBLP:conf/secdev/Rahaman2017} uses static taint analysis to discover library-level cryptographic implementation issues in C/C++ cryptographic libraries (e.g., OpenSSL). It uses symbolic execution based path exploration to reduce false alarms, which is usually costly. 
%
Researchers found that misusing non-cryptographic APIs in Android also have serious security implications. These APIs include APIs to access sensitive information (such as location, IMEI, and passwords)~\cite{DBLP:conf/ndss/NanY0ZZ018}, APIs for fingerprint protection~\cite{DBLP:conf/ndss/BianchiFMKVCL18}, and cloud service APIs for information storage~\cite{zuodoes}. The methodology described in this paper can be applied to address these APIs.
Recently, data driven techniques to identify API misuses have been proposed~\cite{DBLP:conf/pldi/PaletovTRV18, DBLP:conf/icse/0001URRK18}. These techniques uses lightweight static analysis to infer rules from examples that can be used for detection. 
In~\cite{DBLP:conf/sigsoft/MuraliCJ17}, authors proposed a Bayesian framework for automatically learning correct API uses that can be used for anomaly-based API misuse detection.
Efforts on automatically repairing insecure code have also been reported~\cite{DBLP:conf/esorics/MaTLSD17,DBLP:conf/ccs/MaLLD16,DBLP:conf/kbse/PhamNNN10}.
Static code analysis has been extensively used for other related software problems as well, including malware analysis and detection~\cite{DBLP:conf/ndss/PanWDWY17,Karim-CS-2015,Yao-CSF-15}, vulnerability discoveries~\cite{kwon2017a2c, fingeranto}, and data-leak detection~\cite{DBLP:conf/ccs/BosuLYW17}. In~\cite{DBLP:conf/nsdi/ChiCNRS17}, Chi~\etal presented a system to infer client behaviors by leveraging symbolic executions of client-side code. They used such knowledge to filter anomalous traffic. 
Fuzzing has been demonstrated to automatically discovering software vulnerabilities~\cite{DBLP:conf/uss/RuiterP15, DBLP:conf/ccs/Somorovsky16, DBLP:conf/sp/SivakornAPKJ17}. These techniques aim to find input guided vulnerabilities that result in immediately observable behaviors (e.g., triggering program crashes~\cite{DBLP:conf/ccs/Somorovsky16} or anomalous protocol states~\cite{DBLP:conf/uss/RuiterP15, DBLP:conf/sp/SivakornAPKJ17}). It is unclear how to use fuzzing to detect cryptographic vulnerabilities (e.g., predictable IVs/secrets, legacy primitives) that do not exhibit easily observable anomalous behaviors.
%



\section{Conclusions and an Open Problem}

We described our effort of producing a deployment-quality static analysis tool \myname{} to detect cryptographic misuses in Java programs that developers can routinely use. This effort led to significant new technical contributions, including language-specific contextual refinements for FP reduction, on-demand flow-sensitive, context-sensitive, and field-sensitive program slicing, and benchmark comparisons of leading solutions. We also obtained a trove of security insights into Java secure coding practices.
%
An {\bf open research problem} is designing a compiler that automatically transforms a cryptographic vulnerability or rule into a static-analysis-based code-screening algorithm, similar to what CrySL partially provides, but with much higher expressiveness, precision, and recall.



\bibliographystyle{abbrv}
\bibliography{paper}

\section{Appendix}

\subsection{Other Refinement insights}\label{other:refinements}

\noindent
{\bf RI-III: Removal of bookkeeping indices.}

\lstset{language=Java, 
  basicstyle=\ttfamily\footnotesize,
  keywordstyle=\color{blue}\ttfamily,
  stringstyle=\color{red}\ttfamily,
  commentstyle=\color{green}\ttfamily,
  morecomment=[l][\color{magenta}]{\#},
  showstringspaces=false,
  xleftmargin=0pt
}

\begin{lstlisting}
byte[] iv = new byte[] {0x0, 0x0, 0x0, 
			0x0, 0x0, 0x0, 0x0, 0x0}
\end{lstlisting}

Consider the Java statement above. After transforming into jimple representation, this statement looks like the following list of instructions. 

\lstset{language=Java, 
  basicstyle=\ttfamily\footnotesize,
  keywordstyle=\color{blue}\ttfamily,
  stringstyle=\color{red}\ttfamily,
  commentstyle=\color{green}\ttfamily,
  morecomment=[l][\color{magenta}]{\#},
  showstringspaces=false,
  xleftmargin=0pt
}
\begin{lstlisting}
   $r15 = newarray (byte)[8]
   $r15[0] = 0
   $r15[1] = 0
   $r15[2] = 0
   $r15[3] = 0
   $r15[4] = 0
   $r15[5] = 0
   $r15[6] = 0

   $r2 = $r15
\end{lstlisting}

The hard coded size and the indices of an array can be regarded as pseudo-influences. To address this false positives, we discard all the constants that influences an array index. Also, any constant that influences the size or the index parameter of a collection can also be regarded as pseudo-influences. We regard \texttt{List}, \texttt{Set} as collections.

\noindent
\textbf{RI-IV: Removal of contextually incompatible constants.}

Clipping of orthogonal invocations that doesn't appear in an assign statement can also cause false positives. To reduce false alarms further, we also discard some constants constants based on its type and context.
Let's consider, a class named PBEInfo is used to store iteration count and salt and the analysis cannot explore PBEInfo class. A basic use-def analysis will report \texttt{5} as a salt from the following invoke instruction: \texttt{specialinvoke r1.<KeyHolder: void <init>(Integer, String)>(5, "5341453")}. However, a standalone Boolean or Integer constant is unlikely to be used as a key, IV or salt, since their corresponding APIs only allow byte arrays. Also, any hard-coded size parameter (e.g., number of iterations in PBE (Rule 13), key size for insecure asymmetric crypto (Rule 15)) is unlikely to have any type other than \textit{Integer}. Therefore, it is possible to discard some of the pseudo-influences by considering the types of a constant based on its context.

\noindent
\textbf{RI-V: Removal of constants in infeasible paths.}

Some constant initializations are overwritten along the path to the point of interest. Counting such constants with infeasible influences will result in false positives.  Since, empty strings and \texttt{null}s are used for initialization purpose and most often, these initialization are replaced with other values. To avoid false positive for this case, depending on rules and the slicing criteria we discard \texttt{null} and empty strings. For example, \texttt{SecretKeySpec} prohibits keys to be \texttt{null} or empty. \texttt{IvParameterSpec} does not allow \texttt{null} as IV. Also, \texttt{PBEParameterSpec} does not allow the salt to be \texttt{null}.

\subsection{Other Evaluation Results}

\begin{figure}[!hbt]
\small
\centering
\includegraphics[width=0.42\textwidth]{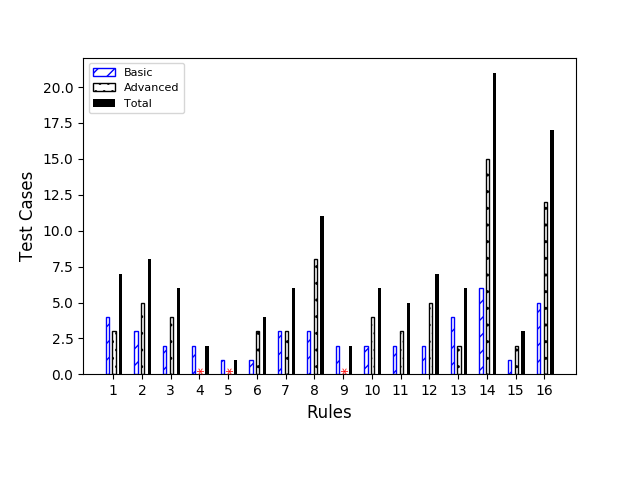}
\vspace{-30pt}
 	\caption{Test cases per Rule in \bench{}.}\label{rules:testcases}
 	\vspace{-15pt}
\end{figure}

\begin{figure}[!hb]
\small
\centering
\includegraphics[width=0.42\textwidth]{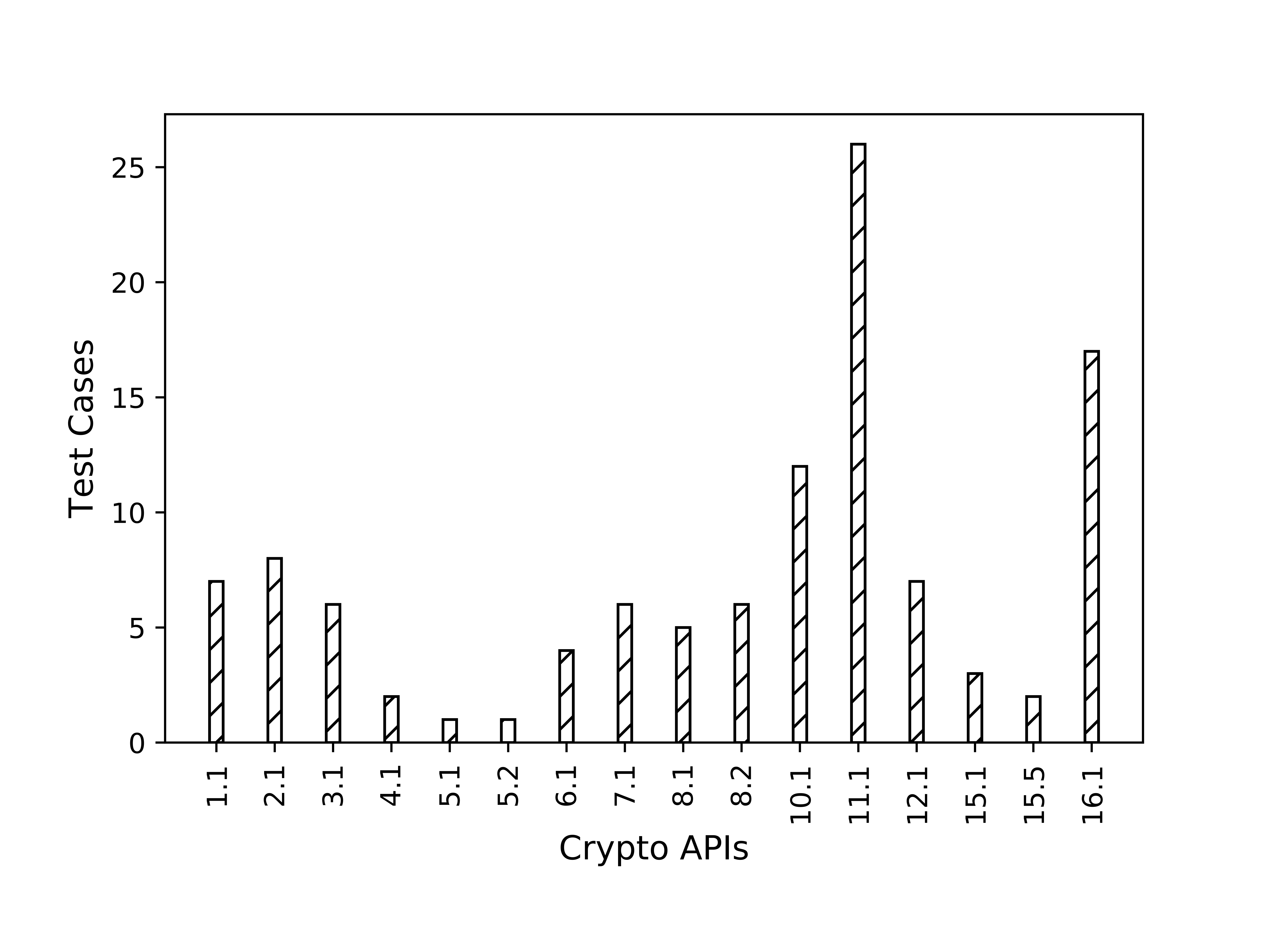}
\vspace{-25pt}
 	\caption{Test cases per API in \bench{}. A test case can cover one ore more APIs (e.g., test cases for Rule 15). APIs corresponding to the labels can be found in Tables~\ref{tab:inter-slicing-criteria},~\ref{tab:methods-to-slice}, and~\ref{tab:forward-slicing-criteria}.}\label{apis:testcases}
\end{figure}

\begin{table*}[!h]
	\centering
	\caption{The number of alerts in Apache (total 94 root-subprojects) and Android applications (6,181). For Rules 1, 2, 3, 8, 10, 12, each constant/predictable value of an array/collection is considered as an individual violation.}
	\label{overview:table}
	    \footnotesize
		\begin{tabular}{lcclcc}
			\Xhline{2\arrayrulewidth}
			\multicolumn{1}{c}{\multirow{2}{*}{\textbf{Rules}}} & \multicolumn{2}{c}{\textbf{Apache}} &  & \multicolumn{2}{c}{\textbf{Android}} \\ \cline{2-3} \cline{5-6} 
			\multicolumn{1}{c}{} & \textbf{\# of Root-subprojects} & \textbf{\# of Alerts Per Rule} &  & \textbf{\# of Applications} & \textbf{\#of Alerts Per Rule} \\ \Xhline{2\arrayrulewidth}
			(1,2) Predictable Keys & 37 (39.36\%) & 264 &  & 1,617 (26.16\%) & 12,457 \\
			(3) Hardcoded Store Password & 29 (30.85\%) & 148 &  & 218 (3.52\%) & 458 \\ \hline
			(4) Dummy Hostname Verifier & 8 (8.51\%) & 12 & \multicolumn{1}{c}{} &  800 (12.94\%) & 1,155 \\
			(5) Dummy Cert. Validation & 11 (11.70\%) & 30 &  & 1,564 (25.30\%) & 3,856 \\
			(6) Used Improper Socket & 4 (4.25\%) & 4 &  & 210 (3.39\%) & 271 \\
			(7) Used HTTP & 24 (29.62\%) & 222 & \multicolumn{1}{c}{} & 2,486 (40.22\%) & 8,321 \\ \hline
			(8) Predictable Seeds & 0 (0\%) & 0 &  & 80 (1.29\%) & 544 \\
			(9) Untrusted PRNG & 33 (35.10\%) & 142 &  & 5,194 (84.03\%) & 36,223\\ \hline
			(10) Static Salts & 21 (22.34\%) & 112 &  & 199 (3.21\%) & 1,757 \\
			(11) ECB mode for Symm. Crypto & 16 (17.02 \%) & 41 &  & 882 (14.26\%) & 1,780 \\
			(12) Static IVs & 4 (4.25 \%) & 41 &  & 913 (14.77\%) & 12,089 \\ \hline
			(13) \textless 1000 PBE Iterations & 25 (26.59 \%) & 43 &  & 151 (2.44\%) & 312 \\
			(14) Broken Symm. Crypto Algorithms & 29 (30.85 \%) & 86 &  & 701 (11.34\%) & 1,742 \\
			(15) Insecure Asymm. Crypto & 9 (10.98 \%) & 12 &  & 108 (1.74\%) & 111 \\ 
			(16) Broken Hash & 42 (44.68 \%) & 138 &  & 5,272 (85.29\%) & 49,769 \\ \Xhline{2\arrayrulewidth}
		\end{tabular}%
	\vspace{20pt}
\end{table*}

\begin{scriptsize}
\begin{table*}[h]
\caption{Benchmark comparison of CrySL, Coverity, SpotBugs, and CryptoGuard on all 16 rules with \bench{}'s 112 test cases. There are 16 secure API use cases (13 in basic and 3 in advanced), which a tool should not raise any alerts on. \myname{} successfully passed these 16 test cases. GTP stands for ground truth positive, which is the number of positives in the benchmark. \myname{} has 11 false negatives, which we reported in Section~\ref{sec:findings} and discussed in Section~\ref{sec:discussion}.}\label{table:totalbench}
\begin{tabular}{|c|l|l|c|c|c|c|c|c|c|c|}
\hline
{No.} & \multicolumn{1}{c|}{\textbf{Rules}} & \multicolumn{1}{c|}{\textbf{GTP}} & \multicolumn{2}{c|}{\textbf{CrySL}} & \multicolumn{2}{c|}{\textbf{Coverity}} & \multicolumn{2}{c|}{\textbf{SpotBugs}} & \multicolumn{2}{c|}{\textbf{CryptoGuard}} \\ \cline{4-11} 
                         & \multicolumn{1}{c|}{} & \multicolumn{1}{c|}{}                       & TP           & FP          & TP             & FP           & TP            & FP            & TP              & FP             \\ \hline
1                        & Predictable Cryptographic Key & 5              & 0            & 4           & 3              & 0            & 2             & 0             & 5               & 0              \\ \hline
2                        & Predictable Password for PBE & 6               & 0            & 2           & 5              & 0            & 3             & 0             & 6               & 0              \\ \hline
3                        & Predictable Password for KeyStore &5          & 0            & 5           & 3              & 0            & 2             & 0             & 5               & 0              \\ \hline
4                        & Dummy Hostname Verifier & 1                    & --           & --          & 1              & 0            & 1             & 0             & 1               & 0              \\ \hline
5                        & Dummy Cert. Validation  & 1                    & --           & --          & 1              & 0            & 1             & 0             & 1               & 0              \\ \hline
6                        & Used Improper Socket & 4                       & --           & --          & 4              & 0            & --            & --            & 4               & 0              \\ \hline
7                        & Use of HTTP  & 4                               & --           & --          & --             & --           & --            & --            & 4               & 0              \\ \hline
8                        & Predictable Seed & 10                            & --           & --          & 1              & 0            & --            & --            & 5               & 0              \\ \hline
9                        & Untrusted PRNG & 1                             & --           & --          & --             & --           & 1             & 0             & 1               & 0              \\ \hline
10                       & Static Salt  & 5                               & 5            & 1           & --             & --           & --            & --            & 3               & 0              \\ \hline
11                       & ECB in Symm. Crypto & 4                         & 2            & 1           & 1              & 0            & 1             & 1             & 4               & 0              \\ \hline
\multicolumn{1}{|l|}{12} & Static IV  & 6                                   & 0            & 6           & --             & --           & 6             & 0             & 4               & 0              \\ \hline
\multicolumn{1}{|l|}{13} & \textless{}1000 PBE Iteration &  5               & 2            & 1           & --             & --           & --            & --            & 4               & 0              \\ \hline
\multicolumn{1}{|l|}{14} & Broken Symm. Crypto  & 20                       & 10           & 5           & 4              & 0            & 5             & 5             & 20              & 0              \\ \hline
\multicolumn{1}{|l|}{15} & Insecure Asymm. Crypto  & 3                    & 2            & 1           & --             & --           & 0             & 1             & 2               & 0              \\ \hline
\multicolumn{1}{|l|}{16} & Broken Hash  & 16                               & 8            & 4           & 4              & 0            & 4             & 4             & 16              & 0              \\ \hline
\multicolumn{2}{|r|}{\textbf{Total}} & \textbf{96}                                   & \textbf{29}  & \textbf{30} & \textbf{27}    & \textbf{0}   & \textbf{26}   & \textbf{11}   & \textbf{85}     & \textbf{0}     \\ \hline
\end{tabular}
\vspace{20pt}
\end{table*}
\end{scriptsize}

\begin{table*}[!ht]
	\caption{Rules that use intra-procedural backward program slicing to slice implemented methods of standard Java APIs and their corresponding slicing criteria.}\label{tab:methods-to-slice}
 	\begin{footnotesize}
 	\begin{tabular}{|>{\xsmall}m{0.3cm}|>{\xsmall}m{12.5cm}|>{\xsmall}m{0.7cm}|>{\xsmall}m{2.2cm}|}
 		\hline
 		No. & Method to Slice & Rule & Criterion \\
 		\hline \hline
	 4.1 & \texttt{javax.net.ssl.HostnameVerifier: boolean verify(String,SSLSession)} & 4 & \texttt{return} \\
	\hline
	5.1 & \texttt{void checkServerTrusted(X509Certificate[],String)} & 5 & \texttt{checkValidity()} \\
	5.2 & \texttt{void checkServerTrusted(X509Certificate[],String)} & 5 & \texttt{throw} \\
	5.3 & \texttt{java.security.cert.X509Certificate[] getAcceptedIssuers()} & 5 & \texttt{return} \\

 	\bottomrule
 	\end{tabular}
 	\end{footnotesize}
 	\vspace{10pt}
\end{table*}

\begin{table*}[!ht]
	\caption{Java APIs used as slicing criteria in our intra-procedural forward program slicing and their corresponding security rules.}\label{tab:forward-slicing-criteria}
 	\begin{footnotesize}
 	\begin{tabular}{|>{\xsmall}m{0.4cm}|>{\xsmall}m{12.3cm}|>{\xsmall}m{0.7cm}|>{\xsmall}m{2.5cm}|}
 		\hline
 		No. & Slicing Criterion for Intra Procedural Forward Program Slicing & Rule & Semantic \\
 		\hline \hline
  	6.1 & \texttt{javax.net.ssl.SSLSocketFactory: SocketFactory getDefault()} & 6 & Create SocketFactory \\
    6.2 & \texttt{javax.net.ssl.SSLContext: SSLSocketFactory getSocketFactory()} & 6 & Create SocketFactory \\
         \hline
    15.1 & \texttt{java.security.KeyPairGenerator: KeyPairGenerator getInstance(java.lang.String)} & 15 & Create KeyPairGenerator \\
    15.2 & \texttt{java.security.KeyPairGenerator: KeyPairGenerator getInstance(String,String)>} & 15 & Create KeyPairGenerator \\
    15.3 & \texttt{java.security.KeyPairGenerator: KeyPairGenerator getInstance(String,Provider)} & 15 & Create KeyPairGenerator \\ 
 
 	\bottomrule
 	\end{tabular}
 	\end{footnotesize}
\end{table*}

\begin{table*}[!ht]
	\caption{Java APIs used as slicing criteria in our inter-procedural backward slicing and their corresponding security rules. Boldface indicates the parameter of interest.}\label{tab:inter-slicing-criteria}
 	\begin{footnotesize}
 	\begin{tabular}{|>{\xsmall}m{0.4cm}|>{\xsmall}m{12.5cm}|>{\xsmall}m{0.8cm}|>{\xsmall}m{2.2cm}|}
 		\hline
 		No. & API & Rule & Semantic \\
 		\hline \hline
      
   1.1 & \texttt{javax.crypto.spec.SecretKeySpec: void <init>(\textbf{byte[]},String)} & 1 & Set key\\
    1.2 & \texttt{javax.crypto.spec.SecretKeySpec: void <init>(\textbf{byte[]},int,int,String)} & 1 & Set key\\
         \hline
    2.1 & \texttt{javax.crypto.spec.PBEKeySpec: void <init>(\textbf{char[]})} & 2 & Set password\\  
    2.2 &  \texttt{javax.crypto.spec.PBEKeySpec: void <init>(\textbf{char[]},byte[],int,int)} & 2 & Set password\\
    2.3 &  \texttt{javax.crypto.spec.PBEKeySpec: void <init>(\textbf{char[]},byte[],int)} & 2 & Set password\\
         \hline
    3.1 & \texttt{java.security.KeyStore: void load(InputStream,\textbf{char[]})} & 3 & Set password\\
    3.2 & \texttt{java.security.KeyStore: void store(OutputStream,\textbf{char[]})} & 3 & Set password\\
    3.3 & \texttt{java.security.KeyStore: void setKeyEntry(String,Key,\textbf{char[]},Certificate[])} & 3 & Set password\\
    3.4 & \texttt{java.security.KeyStore: Key getKey(String,\textbf{char[]})} & 3 & Set password\\
     \hline
    7.1 & 	\texttt{java.net.URL: void <init>(\textbf{String})} & 7 & Set URL \\	
	7.2 & \texttt{java.net.URL: void <init>(\textbf{String},String,String)} & 7 & Set URL \\
	7.3 & \texttt{java.net.URL: void <init>(\textbf{String},String,int,String)} & 7 & Set URL \\
	7.4 & \texttt{okhttp3.Request\$Builder: Request\$Builder url(\textbf{String})} & 7 & Set URL \\
	7.5 & \texttt{retrofit2.Retrofit\$Builder: Retrofit\$Builder baseUrl(\textbf{String})} & 7 & Set URL \\
   \hline
  
   8.1 & \texttt{java.security.SecureRandom: void <init>(\textbf{byte[]})} & 8 & Set seed \\
   8.2 & \texttt{java.security.SecureRandom: void setSeed(\textbf{byte[]})} & 8 & Set seed \\
   8.3 &  \texttt{java.security.SecureRandom: void setSeed(\textbf{long})} & 8 & Set seed \\
         \hline
    10.1 & \texttt{javax.crypto.spec.PBEParameterSpec: void <init>(\textbf{byte[]},int)} & 10 & Set salt\\
    10.2 &  \texttt{javax.crypto.spec.PBEParameterSpec: void <init>(\textbf{byte[]},int,AlgorithmParameterSpec)} & 10 & Set salt\\
    10.3 &  \texttt{javax.crypto.spec.PBEKeySpec: void <init>(char[],\textbf{byte[]},int,int)} & 10 & Set salt\\
    10.4 &  \texttt{javax.crypto.spec.PBEKeySpec: void <init>(char[],\textbf{byte[]},int)} & 10 & Set salt\\
         \hline
    11.1 & \texttt{javax.crypto.Cipher: Cipher getInstance(\textbf{String})} & 11, 14  & Select cipher \\
    11.2 & \texttt{javax.crypto.Cipher: Cipher getInstance(\textbf{String}, String)} & 11, 14 & Select cipher   \\
    11.3 & \texttt{javax.crypto.Cipher: Cipher getInstance(\textbf{String}, Provider)} & 11, 14 & Select cipher \\
         \hline
   12.1 &  \texttt{javax.crypto.spec.IvParameterSpec: void <init>(\textbf{byte[]})} & 12 & Set IV \\
    12.2 &  \texttt{javax.crypto.spec.IvParameterSpec: void <init>(\textbf{byte[]},int,int)} & 12 & Set IV \\
     \hline
     13.1 & \texttt{javax.crypto.spec.PBEParameterSpec: void <init>(byte[],\textbf{int})} & 13 & Set iterations \\
    13.2 & \texttt{javax.crypto.spec.PBEParameterSpec: void <init>(byte[],\textbf{int},AlgorithmParameterSpec)} & 13 & Set iterations\\
    13.3 & \texttt{javax.crypto.spec.PBEKeySpec: void <init>(char[],byte[],\textbf{int},int)} & 13 & Set iterations\\
    13.4 & \texttt{javax.crypto.spec.PBEKeySpec: void <init>(char[],byte[],\textbf{int})} & 13 & Set iterations\\                  
          \hline
    15.1 & \texttt{java.security.KeyPairGenerator: KeyPairGenerator getInstance(\textbf{String})} & 15 & Select generator \\
    15.2 &  \texttt{java.security.KeyPairGenerator: KeyPairGenerator getInstance(\textbf{String},String)>} & 15 & Select generator \\
    15.3 &  \texttt{java.security.KeyPairGenerator: KeyPairGenerator getInstance(\textbf{String},Provider)} & 15 & Select generator \\ 
    15.4 &  \texttt{java.security.KeyPairGenerator: void initialize(\textbf{int})} & 15 & Set key size \\
    15.5 &  \texttt{java.security.KeyPairGenerator: void initialize(\textbf{int},java.security.SecureRandom)} & 15 & Set key size  \\
    15.6 &  \texttt{java.security.KeyPairGenerator: void initialize(\textbf{AlgorithmParameterSpec})} & 15 & Set key size  \\
    15.7 &  \texttt{java.security.KeyPairGenerator: void initialize(\textbf{AlgorithmParameterSpec},SecureRandom)} & 15 & Set key size \\
         \hline
         
    16.1 &  \texttt{java.security.MessageDigest: MessageDigest getInstance(\textbf{String})} & 16 & Select hash \\
	16.2 &	\texttt{java.security.MessageDigest: MessageDigest getInstance(\textbf{String}, String)} & 16 & Select hash \\
	16.3 &	\texttt{java.security.MessageDigest: MessageDigest getInstance(\textbf{String}, Provider)} & 16 & Select hash \\
	\hline
    
 	\bottomrule
 	\end{tabular}
 	\end{footnotesize}
\end{table*}

\end{document}